\newif\iffull
\newif\ifsubmission

\documentclass[titlecompat,tight,twocolumn,10pt]{article}
\usepackage{usenix}

\usepackage{xspace}
\usepackage{graphicx}
\usepackage[hyphens]{url}
\usepackage{enumitem}
\usepackage{booktabs}
\usepackage{multirow}
\usepackage[binary-units=true]{siunitx}
\usepackage{amsfonts,amsthm,amsmath}
\usepackage{subcaption}
\usepackage{color}
\usepackage{balance}
\dblfloatsep=\floatsep
\dbltextfloatsep=\textfloatsep
\DeclareMathAlphabet{\mathsc}{OT1}{cmr}{m}{sc}
\newtheorem{theorem}{Theorem}

\theoremstyle{definition}
\newtheorem{heuristic}[theorem]{Heuristic}

\newcommand\sarah[1]{}
\newcommand\haaroon[1]{}
\newcommand\mary[1]{}
\newcommand\george[1]{}

\newcommand*\dash{\ifvmode\quitvmode\else\unskip\kern.16667em\fi---%
\hskip.16667em\relax}

\newenvironment{sarahlist}{
\begin{description}[itemsep=2pt,leftmargin=0.4cm]
}{\end{description}}

\newcommand{\ttot}{t-to-t\xspace}
\newcommand{\ttoz}{t-to-z\xspace}
\newcommand{\ztot}{z-to-t\xspace}
\newcommand{\ztoz}{z-to-z\xspace}
\newcommand{\vjoinsplit}{\textsf{vJoinSplit}\xspace}
\newcommand{\vjoinsplits}{\textsf{vJoinSplits}\xspace}
\newcommand{\vin}{\textsf{zIn}\xspace}
\newcommand{\vout}{\textsf{zOut}\xspace}
\newcommand{\realvin}{\textsf{vIn}\xspace}
\newcommand{\realvout}{\textsf{vOut}\xspace}

\begin{document}

\title{An Empirical Analysis of Anonymity in Zcash}

\ifsubmission\else{
\author{George Kappos, Haaroon Yousaf, Mary Maller, and Sarah
Meiklejohn\\
University College London\\
\texttt{\{georgios.kappos.16,h.yousaf,mary.maller.15,s.meiklejohn\}@ucl.ac.uk}
}
}\fi

\maketitle

\begin{abstract}

Among the now numerous alternative cryptocurrencies derived from Bitcoin,
Zcash is often touted as the one with the strongest anonymity guarantees, due
to its basis in well-regarded cryptographic research.  In this paper, we examine 
the extent to which anonymity is achieved in the deployed version of Zcash. 
We investigate all facets of anonymity 
in Zcash's transactions, ranging from its transparent transactions 
to the interactions with and within its main privacy feature, 
a \emph{shielded pool} that acts as the anonymity set for users 
wishing to spend coins privately.  We conclude that while it is possible to 
use Zcash in a private way, it is also possible to shrink its anonymity set 
considerably by developing simple heuristics based on identifiable patterns 
of usage.

\end{abstract}

\section{Introduction}

Since the introduction of Bitcoin in 2008~\cite{satoshi-bitcoin}, 
cryptocurrencies have become increasingly popular to the point of reaching 
a near-mania, with thousands of deployed 
cryptocurrencies now collectively attracting trillions of dollars in
investment.  While the broader positive potential of ``blockchain'' (i.e., 
the public decentralized ledger underlying almost all cryptocurrencies) is 
still unclear, despite the growing number of legitimate users there are today
still many people using these cryptocurrencies for
less legitimate purposes.  These range from the purchase of drugs or other
illicit goods on so-called dark markets such as Dream Market, to the payments
from victims in ransomware attacks such as WannaCry, with many other 
crimes in between.  Criminals engaged in
these activities may be drawn to Bitcoin due to the relatively low friction 
of making international payments using only pseudonyms as identifiers, but 
the public nature of its ledger of transactions raises the question of how 
much anonymity is actually being achieved.

Indeed, a long line of research~\cite{reid2013analysis,FC:RonSha13,FC:AKRSC13,sarah-fistfulofbitcoins,FC:SpaMagZan14}
has by now demonstrated that the use of pseudonymous addresses in Bitcoin
does not provide any meaningful level of anonymity.  Beyond academic research,
companies now provide analysis of the Bitcoin blockchain as a 
business~\cite{elliptic-article}.  
This type
of analysis was used in several arrests associated with the
takedown of Silk Road~\cite{silk-road}, and to identify 
the attempts of the WannaCry hackers to move their ransom earnings from
Bitcoin into Monero~\cite{wannacry}.

Perhaps in response to this growing awareness that most cryptocurrencies do
not have strong anonymity guarantees, a number of alternative
cryptocurrencies or other privacy-enhancing techniques have been deployed with
the goal of improving on these guarantees.  The most notable cryptocurrencies
that fall into this former category are Dash~\cite{dash} (launched in January
2014), Monero~\cite{monero} (April 2014), and Zcash~\cite{zcash} 
(October 2016).  At the time of this
writing all have a market capitalization of over 1 billion
USD~\cite{coinmarketcap}, although this
figure is notoriously volatile, so should be taken with a grain of salt.

Even within this category of privacy-enhanced cryptocurrencies, and despite
its relative youth, Zcash stands
somewhat on its own.  From an academic perspective, Zcash is backed by highly
regarded research~\cite{SP:MGGR13,SP:BCGGMT14}, and thus comes with seemingly
strong anonymity guarantees.  Indeed, the original papers cryptographically
prove the security of the main privacy feature of Zcash (known as the 
\emph{shielded pool}), in which users can spend shielded coins without 
revealing which coins
they have spent.  
These strong guarantees have attracted at least some criminal 
attention to Zcash: the underground
marketplace AlphaBay was on the verge of accepting it before their shutdown 
in July 2017~\cite{alphabay}, and the Shadow Brokers hacking group started
accepting Zcash in May 2017 (and in fact for their monthly dumps accepted 
exclusively Zcash in September 2017)~\cite{tsb-zcash}.

Despite these theoretical privacy guarantees, the deployed version of
Zcash does not require all transactions to take place within the shielded pool
itself: it also supports so-called \emph{transparent} transactions, which are
essentially the same as transactions in Bitcoin in that they reveal the
pseudonymous addresses of both the senders and recipients, and the amount 
being sent.  It
does require, however, that all newly generated coins pass through the 
shielded pool before being spent further, thus ensuring that all coins have 
been shielded at least once.  This requirement led the Zcash developers to 
conclude that the anonymity set for users spending shielded coins is in fact
all generated coins, and thus that 
``the mixing strategies that other cryptocurrencies use for anonymity 
provide a rather small [anonymity set] in comparison to Zcash'' and 
that ``Zcash has a distinct advantage in terms of transaction 
privacy''~\cite{zcash-faqs}. 

In this paper, we provide the first in-depth empirical analysis of anonymity
in Zcash, in order to examine these claims and more generally provide a
longitudinal study of how Zcash has evolved and who its main
participants are.
We begin in Section~\ref{sec:zcash-usage} by providing a
general examination of the Zcash blockchain, from which we observe that the 
vast majority of Zcash activity is in the transparent part of the 
blockchain, meaning it does not engage with the shielded pool at all.  In
Section~\ref{sec:T-to-T}, we explore this aspect of Zcash by adapting the 
analysis that has already been developed for Bitcoin, and find that exchanges 
typically dominate this part of the blockchain.

We then move in Section~\ref{sec:pool-interaction} to examining interactions
with the shielded pool.  We find that, unsurprisingly, the main actors
doing so are the founders and miners, who are required to 
put all newly generated coins directly into it.  Using newly
developed heuristics for attributing transactions to founders and miners, we
find that 65.6\% of the value withdrawn from the pool can be
linked back to deposits made by either founders or miners.  We also implement
a general heuristic for linking together other types of transactions, and
capture an additional 3.5\% of the value using this.  Our relatively simple
heuristics thus reduce the size of the overall anonymity set by 69.1\%.

In Section~\ref{sec:z-to-z}, we then look at the relatively small percentage
of transactions that have taken place within the shielded pool.  Here, we find
(perhaps unsurprisingly) that relatively little information can be inferred,
although we do identify certain patterns that may warrant further
investigation.  Finally, we perform a small case study of the activities of
the Shadow Brokers within Zcash in Section~\ref{sec:shadow-brokers}, and in
Section~\ref{sec:conclusions} we conclude.

All of our results have been disclosed, at the time of the paper's 
submission, to the creators of Zcash, and discussed extensively with them 
since.  This has resulted in changes to both their public 
communication about Zcash's anonymity as well as the transactional behavior 
of the founders.  Additionally, all the code for our analysis is available as
an open-source
repository.\footnote{{\scriptsize
\url{https://github.com/manganese/zcash-empirical-analysis}}}

\section{Related work}\label{sec:related}

We consider as related all work that has focused on the anonymity of
cryptocurrencies, either by building solutions to achieve stronger
anonymity guarantees or by demonstrating its limits.

In terms of the former, there has been a significant volume of research in
providing solutions for existing cryptocurrencies that allow
interested users to mix their coins in a way that achieves better anonymity
than regular
transactions~\cite{FC:BNMCKF14,FCW:ValRow15,tumblebit-ndss,coinjoin,ESORICs:RufMorKat14,bissias2014sybil,SP:KMSWP16,moebius}.  
Another line of research has focused on producing alternative privacy-enhanced 
cryptocurrencies.  Most notably, Dash~\cite{dash} 
incorporates the techniques of CoinJoin~\cite{coinjoin} in its PrivateSpend 
transactions; Monero~\cite{monero,noether2016ring} uses ring signatures to 
allow users to create ``mix-ins'' (i.e., include the keys of other users in
their own transactions as a way of providing a larger anonymity set); and
Zcash~\cite{zcash,SP:BCGGMT14} uses
zero-knowledge proofs to allow users to spend coins without revealing which
coins are being spent. 

In terms of the latter, there has also been a significant volume of research
on de-anonymizing
Bitcoin~\cite{reid2013analysis,FC:RonSha13,FC:AKRSC13,sarah-fistfulofbitcoins,FC:SpaMagZan14}.  Almost all of these attacks follow the same pattern: they first 
apply so-called clustering heuristics that associate multiple different addresses
with one single entity, based on some evidence of shared ownership.  The
most common assumption is that all input addresses in a transaction belong to 
the same entity, with some papers~\cite{FC:AKRSC13,sarah-fistfulofbitcoins} also
incorporating an additional heuristic in which output addresses receiving 
change are also linked.  Once these clusters are 
formed, a ``re-identification attack''~\cite{sarah-fistfulofbitcoins} then
tags specific addresses and thus the clusters in which they are contained.
These techniques have also been applied to alternative cryptocurrencies with
similar types of transactions, such as Ripple~\cite{moreno2016listening}.

The work that is perhaps closest to our own focuses on de-anonymizing the
privacy solutions described above, rather than just on Bitcoin.  Here, several
papers have focused on analyzing so-called privacy overlays or mixing services 
for Bitcoin~\cite{moeser-ecrime,FCW:MeiOrl15,moeser-join,moeser-anon}, and
considered both their level of anonymity and the extent to which participants
must trust each other.  Some of this analysis~\cite{moeser-anon,FCW:MeiOrl15} 
also has implications for anonymity in Dash, due to its focus on CoinJoin.  More
recently, Miller et al.~\cite{miller2017empirical} and Kumar et
al.~\cite{esorics2017} looked at Monero.  They both found that it was possible 
to link together transactions based on temporal patterns, and also based on
certain patterns of usage, such as users who choose to do transactions with 0 
mix-ins (in which case their ring signature provides no anonymity, which in 
turns affects other users who may have included their key in their own mix-ins).  
Finally, we are aware
of one effort to de-anonymize Zcash, by Quesnelle~\cite{zcash-anon}.  
This article focuses on linking together the
transactions used to shield and deshield coins, based on their timing and the 
amount sent in the transactions.  In comparison, our paper implements this
heuristic but also provides a broader
perspective on the entire Zcash ecosystem, as well as a more in-depth analysis
of all interactions with (and within) the shielded pool.

\section{Background}\label{sec:back}

\subsection{How Zcash works}\label{sec:back-zcash}

Zcash (ZEC) is an alternative cryptocurrency developed as a (code) fork of 
Bitcoin that
aims to break the link between senders and recipients in a transaction.  In
Bitcoin, recipients receive funds into addresses (referred to as the \realvout
in a transaction), and when they spend them they do so from these addresses
(referred to as the \realvin in a transaction).  The act of spending bitcoins 
thus creates a
link between the sender and recipient, and these links can be followed as
bitcoins continue to change hands.  It is thus possible to track any given 
bitcoin from its creation to its current owner.

\begin{figure}[t]
\centering
\includegraphics[width=0.8\linewidth]{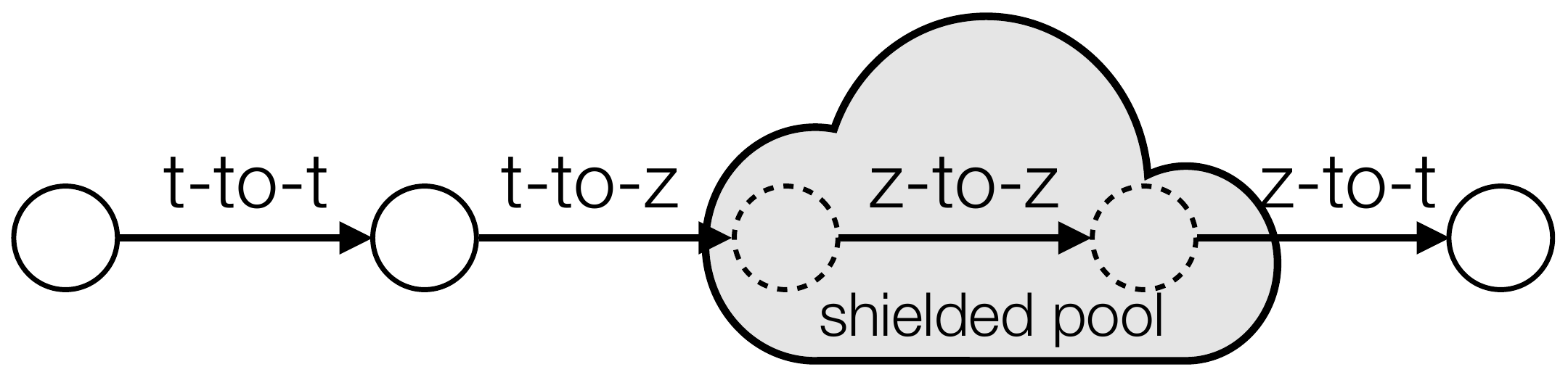}
\caption{A simple diagram illustrating the different types of Zcash 
transactions.  All transaction types are depicted and described with
respect to a single input and output, but can be generalized to handle
multiple inputs and outputs.  In a \ttot transaction, visible quantities of 
ZEC move between visible t-addresses (\vin,\vout$\neq\emptyset$).  In a 
\ttoz transaction, a visible amount of ZEC moves from a visible t-address into 
the shielded pool, at which point it belongs to a hidden z-address
(\vout$=\emptyset$).  In a \ztoz transaction, a hidden
quantity of ZEC moves between hidden z-addresses (\vin,\vout$=\emptyset$).  
Finally, in a \ztot transaction, a hidden quantity of ZEC moves from a hidden 
z-address out of the shielded pool, at which point a visible quantity of it 
belongs to a visible t-address (\vin$=\emptyset$).}
\label{fig:diagram}
\end{figure}

Any transaction which interacts with the so-called shielded pool in Zcash does 
so through the inclusion of a \emph{\vjoinsplit}, which specifies where the
coins are coming from and where they are going.  To 
receive funds, users can provide either a transparent address (t-address) or 
a shielded address (z-address).  Coins that are held in z-addresses are said
to be in the shielded pool.

To specify where the funds are going, a \vjoinsplit contains (1) a list 
of output t-addresses with funds assigned to them (called
\emph{\vout}), (2) two shielded outputs, and (3) an encrypted memo field.  
The \vout can be empty, in which case the transaction is either
\emph{shielded} (\ttoz) or \emph{private} (\ztoz), depending on the inputs.
If the \vout list contains a quantity of ZEC not assigned to any address, then
we still consider it to be empty (as this is simply the allocation of the
miner's fee).  Each shielded output contains an unknown quantity of ZEC as
well as a hidden double-spending token.  The shielded output can be a dummy
output (i.e., it contains zero ZEC) to hide the fact that there is no shielded
output.  The encrypted memo field can be used to send private messages to the
recipients of the shielded outputs.

To specify where the funds are coming from, a \vjoinsplit also contains (1) a 
list of input t-addresses (called \emph{\vin}), (2) two double-spending tokens, 
and (3) a zero-knowledge proof.  The \vin can be empty, in
which case the transaction is either \emph{deshielded} (\ztot) if \vout is not
empty, or private (\ztoz) if it is.  Each double-spending token is either a 
unique token belonging to some previous
shielded output, or a dummy value used to hide the fact that there is no
shielded input.  The double-spending token does not reveal to which shielded
output it belongs.  The zero-knowledge proof guarantees two things.  First, it
proves that
the double-spending token genuinely belongs to some previous shielded output.
Second, it proves that the sum of (1) the values in the addresses in \vin plus 
(2) the values 
represented by the double-spending tokens is equal to the sum of (1) the values 
assigned to the addresses in \vout plus (2) the values in the 
shielded outputs plus (3) the miner's fee.  A summary of the different types 
of transactions is in Figure~\ref{fig:diagram}.

%

\subsection{Participants in the Zcash ecosystem}\label{sec:back-participants}

In this section we describe four types of participants who interact in the 
Zcash network.

Founders took part in the initial creation and release of Zcash, and
will receive 20\% of all newly generated coins (currently 2.5~ZEC out of the
12.5~ZEC block reward).  The founder addresses are specified in the Zcash 
chain parameters~\cite{params}.

Miners take part in the maintenance of the ledger, and in
doing so receive newly generated coins (10 out of the 12.5~ZEC block reward), 
as well as any fees from the transactions
included in the blocks they mine.  Many miners choose not to mine on their
own, but join a mining pool; a list of mining pools can be found in 
Table~\ref{tab:miners}.  One or many miners win each block,
and the first transaction in the block is a \emph{coin generation} (coingen) 
that assigns newly generated coins to their address(es), as well as to the 
address(es) of the founders.

Services are entities that accept ZEC as some form of payment.  These include
exchanges like Bitfinex, which allow users to trade fiat
currencies and other cryptocurrencies for ZEC (and vice versa), and platforms 
like ShapeShift~\cite{shapeshift}, which allow users to trade within
cryptocurrencies and other digital assets without requiring registration.  

Finally, users are participants who hold and transact in ZEC at a more
individual level.  In addition to regular individuals, this category includes
charities and other organizations that may choose to accept donations in
Zcash.  A notable user is the Shadow Brokers, a hacker group who have 
published several leaks containing hacking tools from the NSA and accept 
payment in Zcash.  We explore their usage of Zcash in 
Section~\ref{sec:shadow-brokers}.

\section{General Blockchain Statistics}\label{sec:zcash-usage}

We used the \texttt{zcashd} client to download the Zcash blockchain, and
loaded a database representation of it into Apache Spark.  We then performed 
our analysis using a custom set of Python scripts equipped with \mbox{PySpark}.  
We last parsed the block chain on January 21 2018, at which point 258,472 blocks 
had been mined.  Overall, 3,106,643~ZEC had been generated since the genesis 
block, out of which 2,485,461~ZEC went to the miners and the rest 
(621,182~ZEC) went to the founders. 

\subsection{Transactions}\label{sec:tx-usage}

\begin{figure}[t]
\centering
\includegraphics[width=0.9\linewidth]{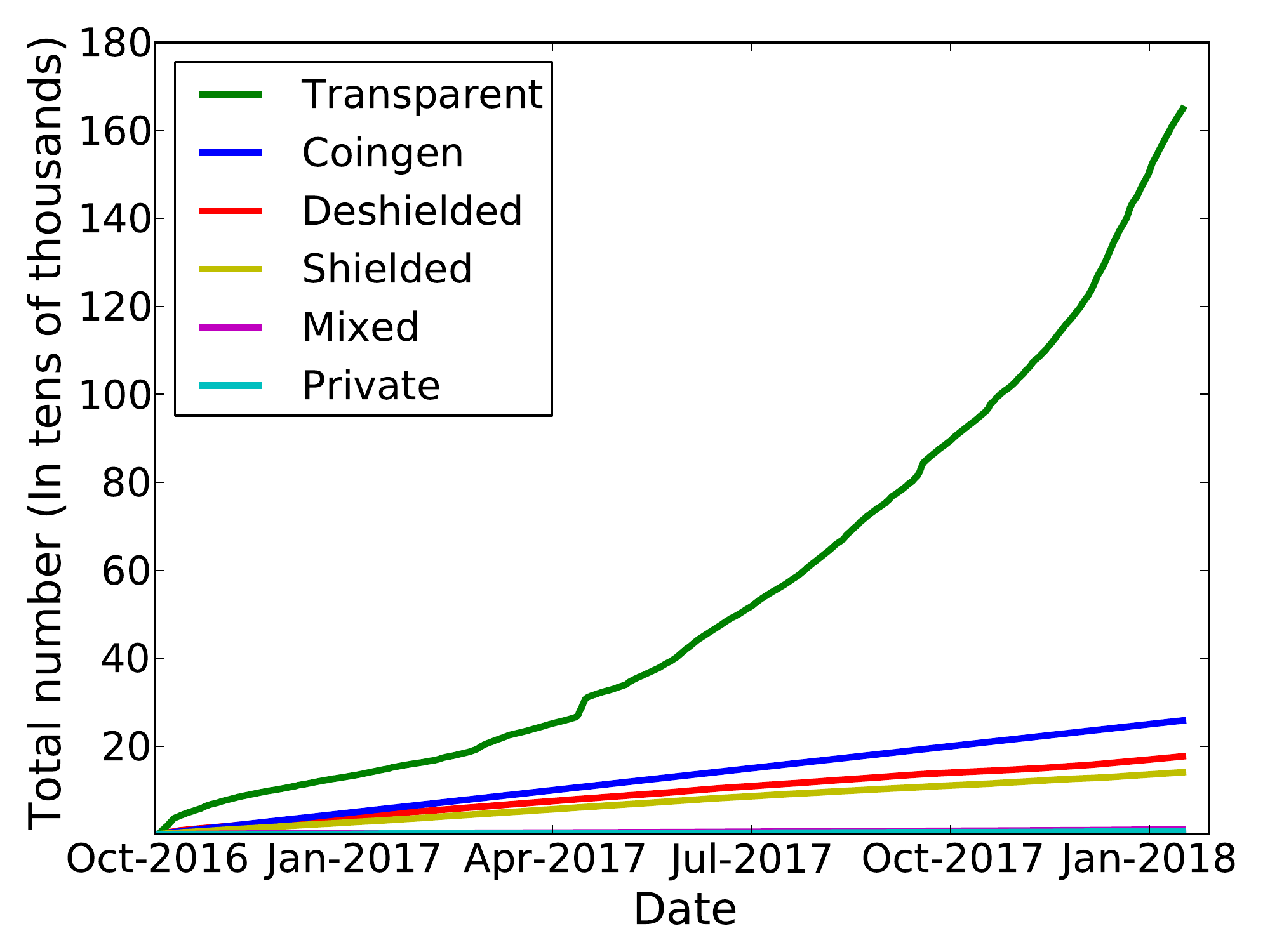}
\caption{The total number of each of the different types of transactions 
over time.}
\label{fig:cumulative_types_of_transactions}
\end{figure}

\begin{figure}[t]
\centering
\includegraphics[width=0.95\linewidth]{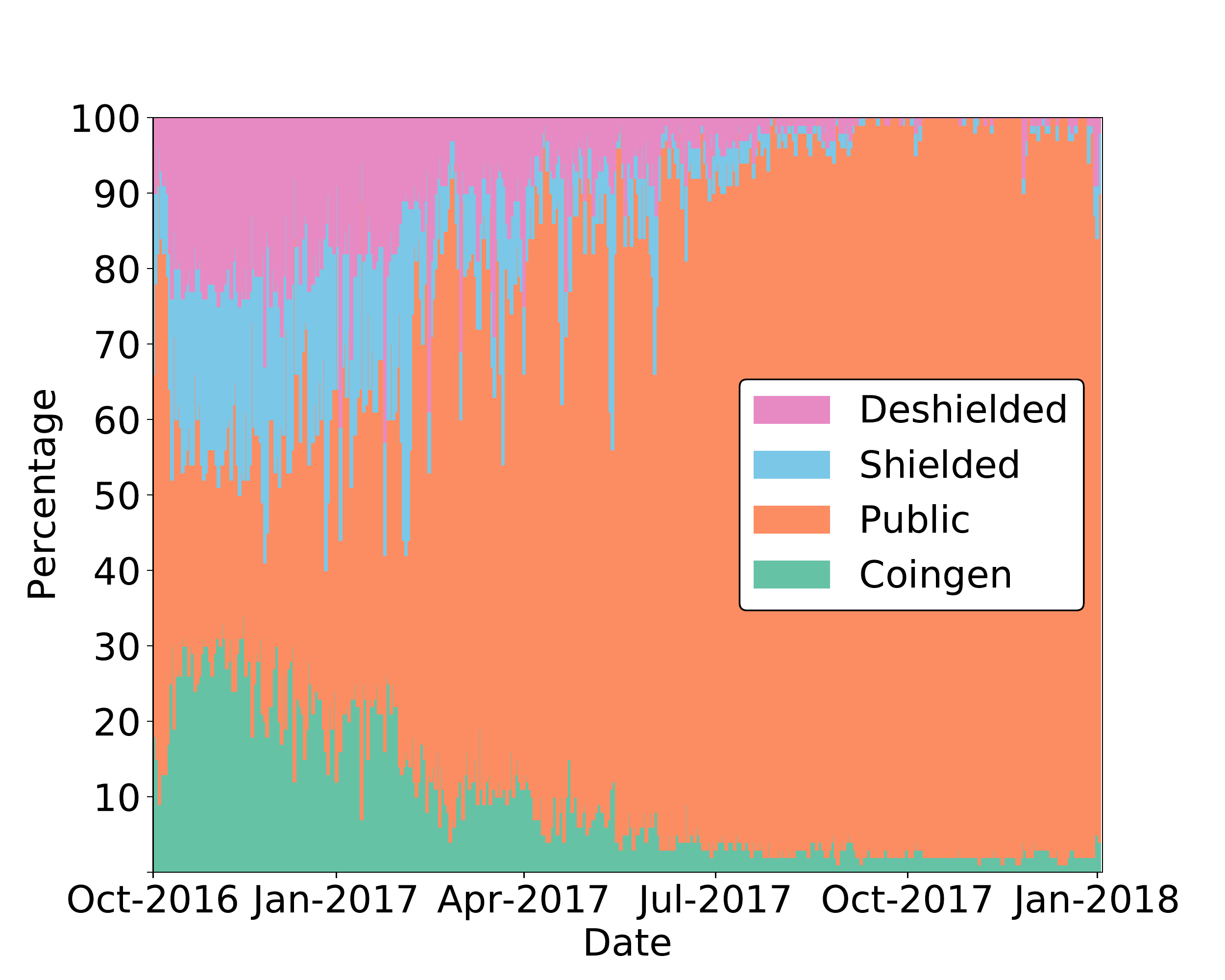}
\caption{The fraction of the value in each block representing each different 
type of transaction over time, averaged daily.  Here, `public' captures both 
transparent transactions and the visible components of mixed transactions.}
\label{fig:fistful_DailyAverage}
\end{figure}

Across all blocks, there were 2,242,847 transactions.  A complete 
breakdown of the transaction types is in Table~\ref{tab:tx-types},
and graphs depicting the growth of each transaction type over time are in
Figures~\ref{fig:cumulative_types_of_transactions}
and~\ref{fig:fistful_DailyAverage}.\footnote{We use the term `mixed' to mean
    transactions that have both a \realvin and a \realvout, and a \vjoinsplit.}    
The vast majority of transactions are public (i.e., either transparent or a 
coin generation).
Of the transactions that do interact with the pool (335,630, or 14.96\%, in
total), only a very small percentage are private transactions; i.e., 
transactions within the pool.
Looking at the types of transactions over time
in Figure~\ref{fig:cumulative_types_of_transactions}, we can see that the
number of coingen, shielded, and deshielded transactions all grow in an
approximately linear fashion.  As we explore 
in Section~\ref{sec:miners}, this correlation is due largely to the
habits of the miners.  Looking at both this figure and
Figure~\ref{fig:fistful_DailyAverage}, we can see that while the number of
transactions interacting with the pool has grown in a relatively linear
fashion, the value they carry has over time become a very small 
percentage of all
blocks, as more mainstream (and thus transparent) usage of Zcash has
increased.

\begin{table}
\centering
\begin{tabular}{lS[table-format=7.0]S[table-format=2.1]}
\toprule
Type & {Number} & {Percentage} \\
\midrule
Transparent & 1648745 & 73.5\\
Coingen & 258472 & 11.5 \\
Deshielded & 177009 & 7.9\\
Shielded & 140796 & 6.3 \\
Mixed & 10891 & 0.5 \\
Private & 6934 & 0.3 \\
\bottomrule
\end{tabular}
\caption{The total number of each transaction type.}
\label{tab:tx-types}
\end{table}


\subsection{Addresses}

Across all transactions, there have been 1,740,378 distinct t-addresses used.
Of these, 8,727 have ever acted as inputs in a \ttoz transaction and 330,780 
have ever acted as outputs in a \ztot transaction.  As we explore in
Section~\ref{sec:miners}, much of this asymmetry is due to the behavior of 
mining pools, which use a small number of addresses to collect the block
reward, but a large number of addresses (representing all the individual
miners) to pay out of the pool.  Given the nature of the shielded pool, it is 
not possible to know the total number of z-addresses used.

\begin{figure}[t]
\centering
\includegraphics[width=0.9\linewidth]{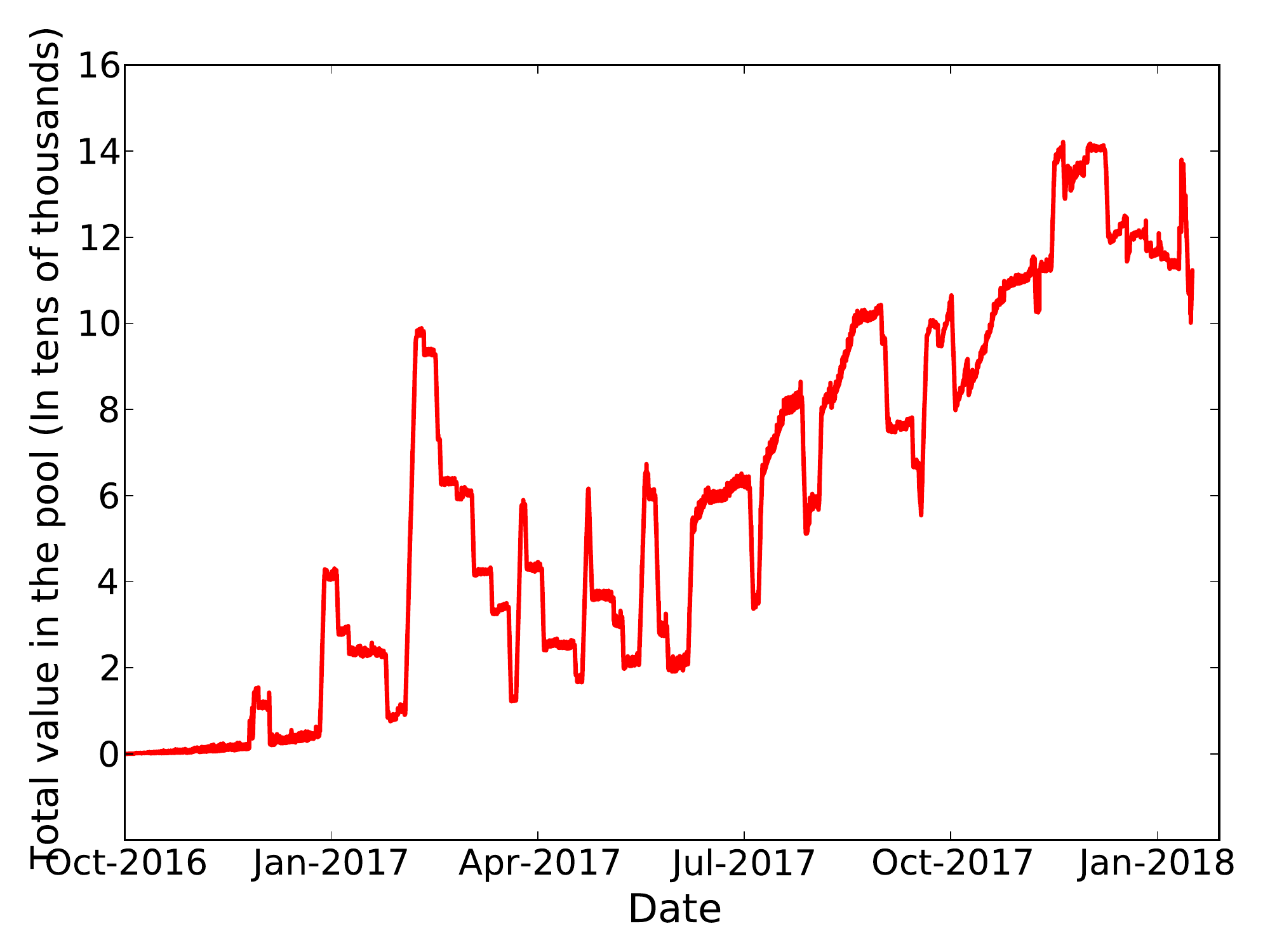}
\caption{The total value in the shielded pool over time.}
\label{fig:total_value_in_pool}
\end{figure}

Figure~\ref{fig:total_value_in_pool} shows the total value in the pool over 
time.  Although the overall value is increasing over time, 
there are certain shielding and de-shielding patterns that create spikes.  As
we explore in Section~\ref{sec:pool-interaction}, these spikes are due largely
to the habits of the miners and founders.  At the time of writing, there are 
112,235 ZEC in the pool, or 3.6\% of the total monetary supply. 

If we rank addresses by their wealth, we first observe that only 25\% of all
t-addresses have a non-zero balance.  Of these, the top 1\% hold 78\% of all
ZEC.  The address with the highest balance had 118,257.75~ZEC, which means
the richest address has a higher balance than the entire shielded pool.

\section{T-Address Clustering} \label{sec:T-to-T}

As discussed in Section~\ref{sec:zcash-usage}, a large proportion of the
activity on Zcash does not use the shielded pool.  This means it is
essentially identical to Bitcoin, and thus can be de-anonymized using the same
techniques discussed for Bitcoin in Section~\ref{sec:related}.  

\subsection{Clustering addresses}\label{sec:clustering}

To identify the usage of transparent addresses, we begin by recalling the 
``multi-input'' heuristic for clustering Bitcoin
addresses.  In this heuristic, addresses that are used as inputs to the same 
transaction are assigned to the same cluster.  In Bitcoin, this heuristic can 
be applied to all transactions, as they are all transparent.  In Zcash, we 
perform this clustering as long as there are multiple input t-addresses.

\begin{heuristic}\label{heuristic:inputclusters}
If two or more t-addresses are inputs in the same transaction (whether that
transaction is transparent, shielded, or mixed), then they are controlled by 
the same entity. 
\end{heuristic} 

In terms of false positives, we believe that these are
at least as unlikely for Zcash as they are for Bitcoin, as Zcash is a direct
fork of Bitcoin and the standard client has the same behavior.  In fact, we 
are not aware of any input-mixing techniques like CoinJoin~\cite{coinjoin} for 
Zcash, so could argue that the risk of false positives is even lower than it 
is for Bitcoin.  As this heuristic has already been used extensively in
Bitcoin, we thus believe it to be realistic for use in Zcash.

We implemented this heuristic by defining each t-address as a node in a graph,
and adding an (undirected) edge in the graph between addresses that had been
input to the same transaction.  The connected components of the graph then
formed the clusters, which represent distinct entities controlling potentially
many addresses.  The result was a set of 560,319 clusters, of which 97,539
contained more than a single address.

As in Bitcoin, using just this one heuristic is already quite effective but
does not capture the common usage of \emph{change addresses}, in which a
transaction sends coins to the actual recipient but then also sends any coins
left over in the input back to the sender.  Meiklejohn et
al.~\cite{sarah-fistfulofbitcoins} use in their analysis a heuristic based on
this behavior, but warn that it is somewhat fragile.  Indeed, their heuristic
seems largely dependent on the specific behavior of several large Bitcoin
services, so we chose not to implement it in its full form.  Nevertheless, we
did use a related Zcash-specific heuristic in our case study of the Shadow
Brokers in Section~\ref{sec:shadow-brokers}.

\begin{heuristic}\label{heuristic:others}
If one (or more) address is an input t-address in a \vjoinsplit transaction
and a second address is an output t-address in the same \vjoinsplit
transaction, then if the size of \vout is $1$ (i.e., this is the only
transparent output address), the second address belongs to the same user who 
controls the input addresses. 
\end{heuristic}

To justify this heuristic, we observe that users may not want to deposit
all of the coins in their address when putting coins into the pool, in which 
case they will have to make change.  The only risk of a false positive is if
users are instead sending money to two separate individuals, one using a 
z-address and one using a t-address.  One notable exception to 
this rule is users of the zcash4win wallet.  Here, the address of the wallet 
operator is an output t-address if the user decides to pay the developer fee,
so would produce exactly this type of transaction for users putting money into
the shielded pool.  This address is identifiable, however, so these types of
transactions can be omitted from our analysis.  Nevertheless, due to concerns
about the safety of this heuristic (i.e., its ability to avoid false
positives), we chose not to incorporate it into our general analysis below.
  
%

\subsection{Tagging addresses}\label{sec:tagging}

Having now obtained a set of clusters, we next sought to assign names to them.
To accomplish this, we performed a scaled-down version of
the techniques used by Meiklejohn et
al.~\cite{sarah-fistfulofbitcoins}.  In particular, given that Zcash is still
relatively new, there are not many different types of services that accept
Zcash. We thus restricted ourselves to interacting with exchanges.

We first identified the top ten Zcash exchanges according to 
volume traded~\cite{coinmarketcap}.  We then created an account with each 
exchange and deposited a small quantity of ZEC into it, tagging as we did 
the output t-addresses in the resulting transaction as belonging to the 
exchange.  We 
then withdrew this amount to our own wallet, and again tagged the t-addresses
(this time on the sender side) as belonging to the exchange.  We occasionally
did several deposit transactions if it seemed likely that doing so would tag
more addresses.  Finally, we also interacted
with ShapeShift, which as mentioned in Section~\ref{sec:back-participants}
allows users to move amongst cryptocurrencies without the need to create an
account.  Here we did a single ``shift'' into Zcash and a single shift out.  A
summary of our interactions with all the different exchanges is in
Table~\ref{tab:ourtxs}.

\begin{table}
\centering
\begin{tabular}{lccc}
\toprule
Service & Cluster & \# deposits & \# withdrawals \\ 
\midrule
Binance  &  7      &  1       &   1       \\
Bitfinex &  3      &  4       &   1		  \\
Bithumb  &  14     &  2       &   1       \\
Bittrex  &  1      &  1       &   1       \\
Bit-z    &  30     &  2       &   1       \\
Exmo     &  4      &  2       &   1       \\
HitBTC   &  18     &  1       &   1       \\
Huobi    &  26     &  2       &   1       \\
Kraken   &  12     &  1       &   1       \\
Poloniex &  0      &  1       &   1       \\
\midrule
ShapeShift & 2   &  1  &  1 \\
zcash4win  & 139 &  1  & 2  \\
\bottomrule 
\end{tabular}
\caption{The services we interacted with, the identifier of the cluster they
were associated with after running Heuristic~\ref{heuristic:inputclusters}, 
and the number of deposits and withdrawals we did with them.  The first ten
are exchanges, ShapeShift is an inter-cryptocurrency exchange, and zcash4win
is a Windows-based Zcash client.}
\label{tab:ourtxs} 
\end{table}

Finally, we collected the publicized addresses of the founders~\cite{params}, 
as well as addresses from known mining pools.  For the latter we started by 
scraping the tags of these addresses from the Zchain explorer~\cite{zchain}.
We then validated them against the blocks advertised on some of the websites of 
the mining pools themselves (which we also scraped) to ensure that they were the 
correct tags; i.e., if the recipient of the coingen transaction in a given 
block was tagged as belonging to a given mining pool, then we checked to see
that the block had been advertised on the website of that mining pool. 
We then augmented these sets of addresses with the addresses tagged as 
belonging to founders and miners according to the heuristics developed in
Section~\ref{sec:pool-interaction}.  We present these heuristics in
significantly more detail there, but they resulted in us tagging 123 founder
addresses and 110,918 miner addresses (belonging to a variety of different
pools).

\subsection{Results}

As mentioned in Section~\ref{sec:clustering}, running
Heuristic~\ref{heuristic:inputclusters} resulted in 560,319 clusters, of 
which 97,539 contained more than a single address.  We assigned each cluster a
unique identifier, ordered by the number of addresses in the cluster, so that 
the biggest cluster had identifier 0.
%
%
%
  
\subsubsection{Exchanges and wallets}

As can be seen in Table~\ref{tab:ourtxs}, many of the exchanges are associated
with some of the biggest clusters, with four out of the top five clusters
belonging to popular exchanges.  In general, we found that the top five
clusters accounted for 11.21\% of all transactions.  Identifying exchanges is
important, as it makes it possible to discover where individual users may have
purchased their ZEC.  Given existing and emerging regulations, they are also 
the one type of participant in the Zcash ecosystem that might know the
real-world identity of users.

In many of the exchange clusters, we also identified large fractions of
addresses that had been tagged as miners.  This implies that individual 
miners use the addresses of their exchange accounts to receive their mining 
reward, which might be expected if their goal is to cash out directly.  We
found some, but far fewer, founder addresses at some of the exchanges as well.

Our clustering also reveals that ShapeShift (Cluster~2) is fairly heavily
used: it had received over 1.1M~ZEC in total and sent roughly the same.  Unlike 
the exchanges, its cluster contained a relatively small number of miner 
addresses (54), which fits with its usage as a way to shift money, rather 
than hold it in a wallet.

\subsubsection{Mining pools and founders}

Although mining pools and founders account for a large proportion of the 
activity in Zcash (as we explore in Section~\ref{sec:pool-interaction}), many 
re-use the same small set of addresses frequently, so do not belong to 
large clusters.  For example, Flypool had three single-address clusters while 
Coinotron, coinmine.pl, Slushpool and Nanopool
each had two single-address clusters. (A list of mining pools can be
found in Table~\ref{tab:miners} in Section~\ref{sec:miners}).
%
Of the coins that we
saw sent from clusters associated with mining pools, 99.8\% of it went into
the shielded pool, which further validates both our clustering and tagging
techniques.

\subsubsection{Philanthropists}


Via manual inspection, we identified three large organizations that accept 
Zcash donations: the Internet Archive, \url{torservers.net}, and Wikileaks.   
Of these, \url{torservers.net} accepts payment only via a z-address, so we
cannot identify their transactions (Wikileaks accepts payment via a z-address
too, but also via a t-address).  Of the 31 donations to the Internet Archive
that we were able to identify, which totaled 17.3~ZEC, 9 of them were made
anonymously (i.e., as \ztot transactions).  On the other hand, all of the
20 donations to Wikileak's t-address were made as \ttot transactions. 
None of these belong to clusters, as they have never sent a transaction. 


\section{Interactions with the Shielded Pool}\label{sec:pool-interaction}

What makes Zcash unique is of course not its t-addresses (since these
essentially replicate the functionality of Bitcoin), but its shielded pool.
To that end, this section explores interactions with the pool at its
endpoints, meaning the deposits into (\ttoz) and withdrawals out of the pool 
(\ztot).  We then explore interactions within the pool (\ztoz transactions) in
Section~\ref{sec:z-to-z}.

\begin{figure}
\centering
\includegraphics[width=0.9\linewidth]{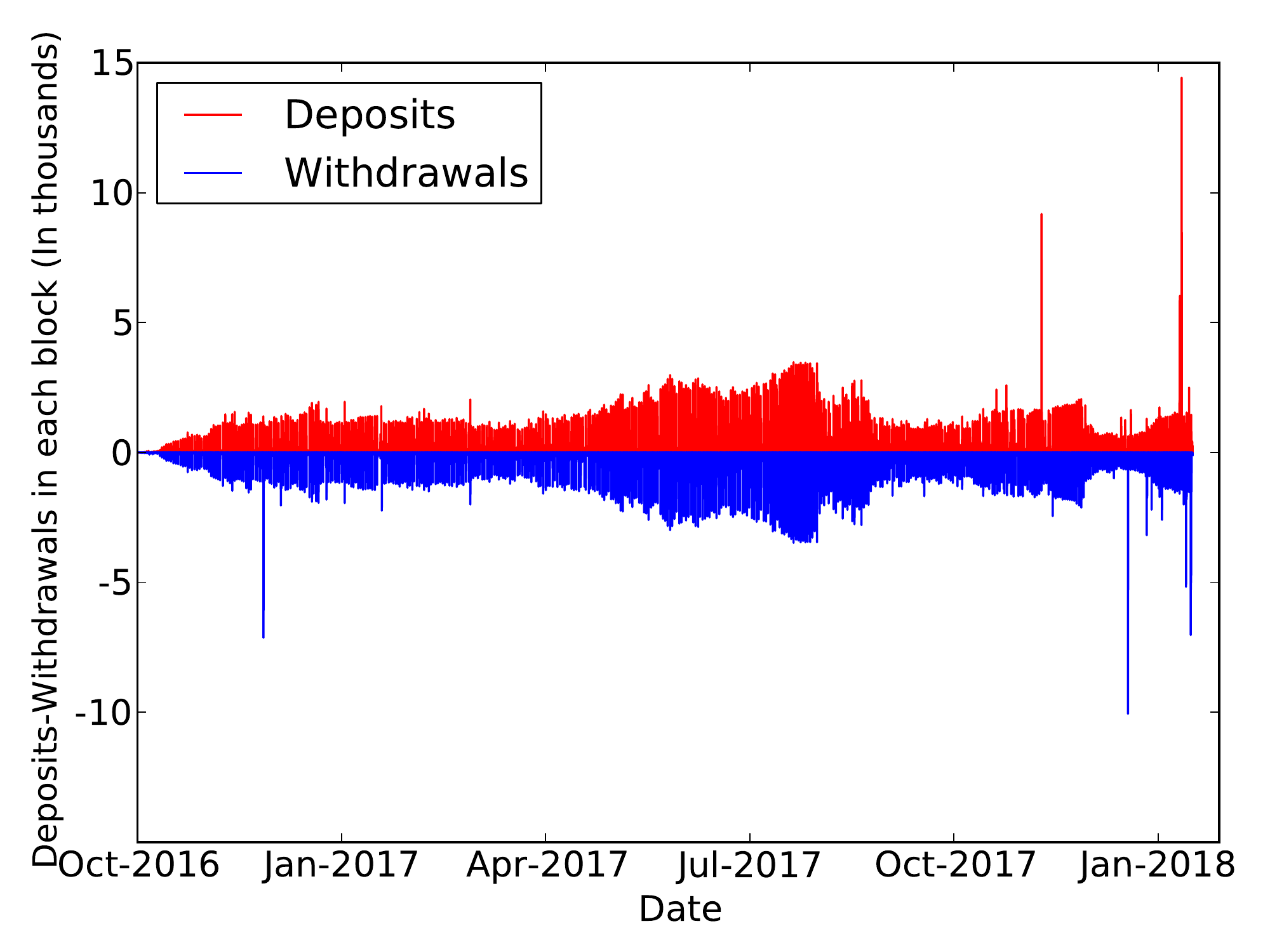}
\caption{Over time, the amount of ZEC put into the shielded pool (in red) and 
the amount taken out of the pool (in blue).}
\label{fig:InAndOutEachBlock}
\end{figure}

To begin, we consider just the amounts put into and taken out of the pool.  
Over time, 3,901,124~ZEC have been deposited into the
pool,\footnote{This is greater than the total number of generated coins, as
    all coins must be deposited into the pool at least once, by the miners or
    founders, but may then go into and out of the pool multiple times.}
and 3,788,889 have been withdrawn.  Figure~\ref{fig:InAndOutEachBlock} plots
both deposits and withdrawals over time.

This figure shows a near-perfect reflection of deposits and
withdrawals, demonstrating that most users not only withdraw the exact number
of ZEC they deposit into the pool, but do so very quickly after the initial
deposit.  As we see in Sections~\ref{sec:founders} and~\ref{sec:miners}, this
phenomenon is accounted for almost fully by the founders and miners.  Looking
further at the figure, we can see that the symmetry is broken occasionally,
and most notably in four ``spikes'': two large withdrawals, and two large
deposits.  Some manual investigation revealed the following:

\begin{sarahlist}
\item[``The early birds''] The first withdrawal spike took place at block
height 30,900, which was created in December 2016.  The cause of the spike was
a single transaction in which 7,135~ZEC was taken out of the pool; given the
exchange rate at that time of 34~USD per ZEC, this was equivalent to
242,590~USD.  The coins were distributed across 15 t-addresses, which initially
we had not tagged as belonging to any named user.  After running the heuristic
described in Section~\ref{sec:founders}, however, we tagged all of these 
addresses as belonging to founders.  In fact, this was the very first
withdrawal that we identified as being associated with founders.
	
\item[``Secret Santa''] The second withdrawal spike took place on December 25
2017, at block height 242,642.  In it, 10,000~ZEC was distributed among 10
different t-addresses, each receiving 1,000~ZEC.  None of these t-addresses 
had done a transaction before then, and none have been involved in one
since (i.e., the coins received in this transaction have not yet been spent).
	
\item[``One-man wolf packs''] Both of the deposit spikes in the graph 
correspond to single large deposits from unknown t-addresses that, 
using our analysis 
from Section~\ref{sec:T-to-T}, we identified as residing in single-address
clusters.  For the first spike, however, many of the deposited amounts came 
directly from a founder address identified by our heuristics 
(Heuristic~\ref{heuristic:founder}), so
given our analysis in Section~\ref{sec:founders} we believe this may also be
associated with the founders.

\end{sarahlist}

While this figure already provides some information about how the pool is used
(namely that most of the money put into it is withdrawn almost immediately
afterwards), it does not tell us who is actually using the pool.  For this, we
attempt to associate addresses with the types of participants identified
in Section~\ref{sec:back-participants}: founders, miners, and `other' 
(encompassing both services and individual users).

\begin{figure}
\centering
\includegraphics[width=0.85\linewidth]{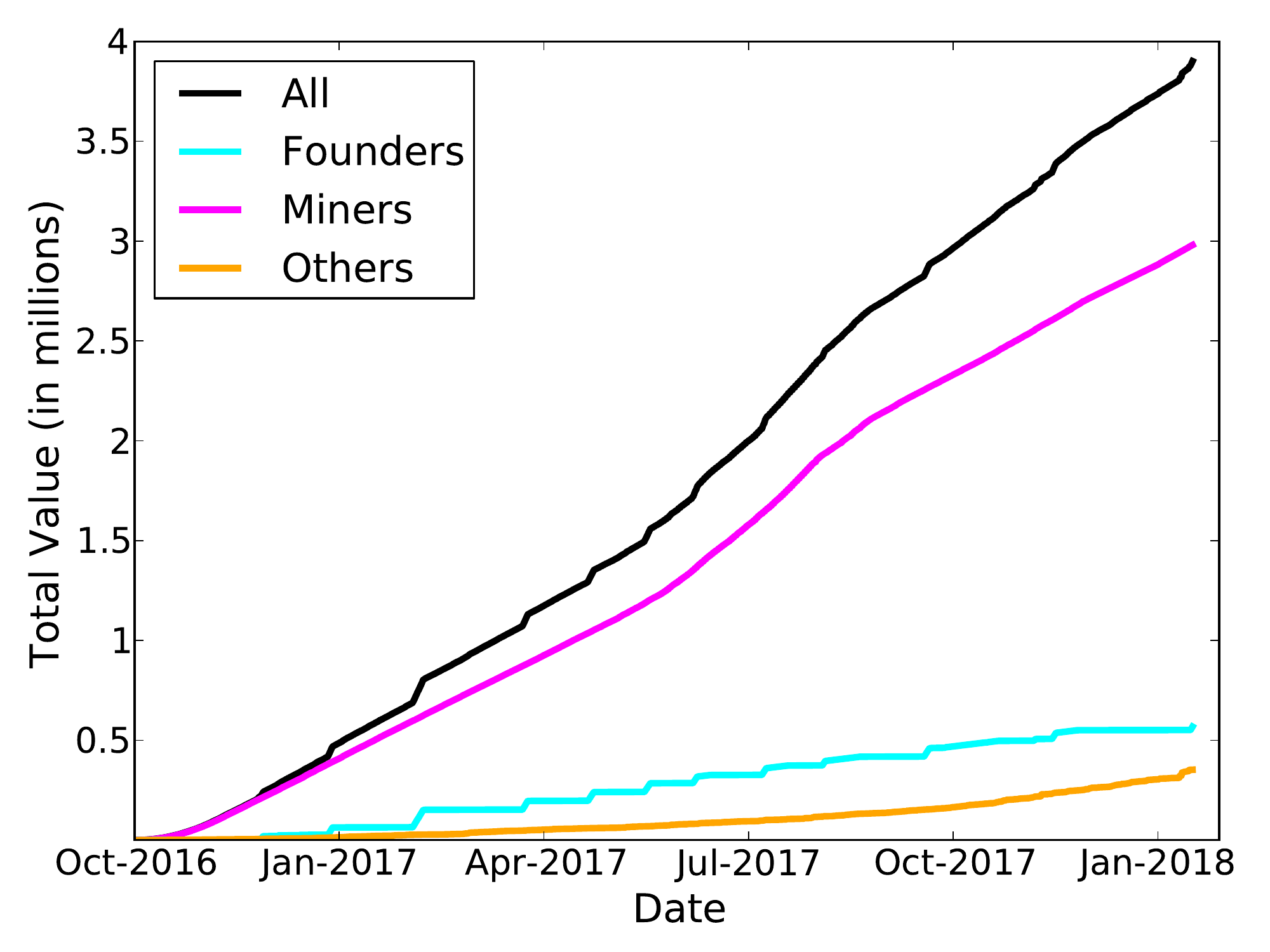}
\caption{Over time, the amount of ZEC deposited into the shielded pool by 
miners, founders, and others.}
\label{fig:t-to-z}
\end{figure}

When considering deposits into the shielded pool, it is easy to 
associate addresses with founders and miners, as the consensus rules dictate 
that they must put their block rewards into the shielded pool before spending 
them further.  
As described in Section~\ref{sec:tagging}, we tagged founders
according to the Zcash parameters, and tagged as miners all recipients of
coingen transactions that were not founders.  We then used these tags to
identify a founder deposit as any \ttoz transaction using one or more
founder addresses as input, and a miner deposit as any \ttoz transaction
using one or more miner addresses as input. 
The results are in Figure~\ref{fig:t-to-z}.

Looking at this figure, it is clear that miners are the main participants
putting money into the pool.  This is not particularly surprising, given that
all the coins they receive must be deposited into the pool at least once, so
if we divide that number of coins by the total number deposited we would
expect at least 63.7\% of the deposits to come from miners. (The actual number
is 76.7\%.) Founders, on the other hand, don't put
as much money into the pool (since they don't have as much to begin with), 
but when they do they put in large amounts that cause visible step-like 
fluctuations to the overall line.  
%

\begin{figure}
\centering
\includegraphics[width=0.8\linewidth]{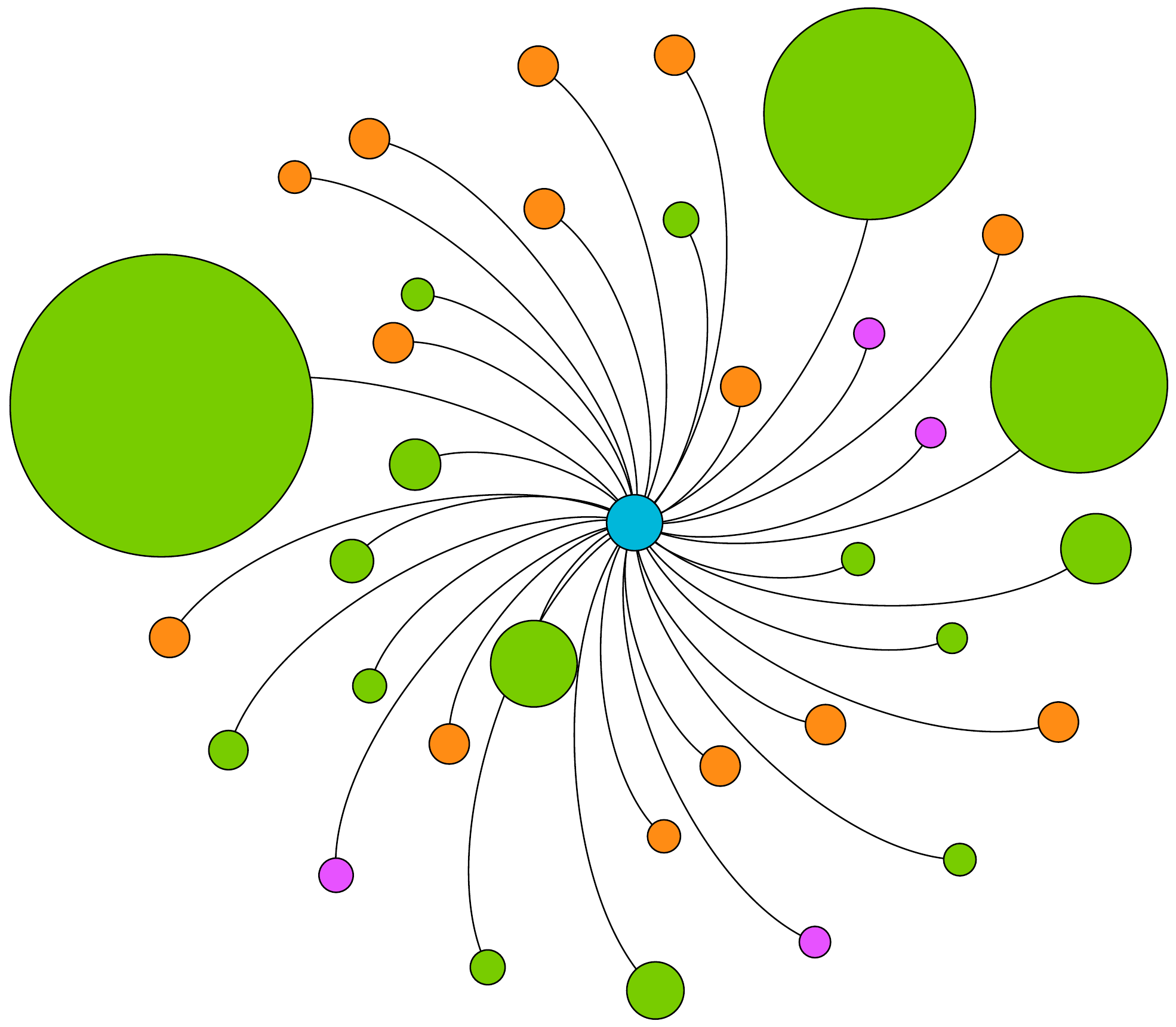}
\caption{The addresses that have put more than 10,000 ZEC into the shielded 
pool over time, where the size of each node is proportional to the
value it has put into the pool.  The addresses of miners are green, of 
founders are orange, and of unknown `other' participants are purple.}
\label{fig:TtoZ10k}
\end{figure}

In terms of the heaviest users, we looked at the individual addresses
that had put more than 10,000 ZEC into the pool.  The results are in
Figure~\ref{fig:TtoZ10k}.

In fact, this figure incorporates the heuristics we develop in
Sections~\ref{sec:founders} and~\ref{sec:miners}, although it looked very
similar when we ran it before applying our heuristics (which makes sense,
since our heuristics mainly act to link \ztot transactions).
Nevertheless, it demonstrates again that most of the heavy users of the pool
are miners, with founders also depositing large amounts but spreading them 
over a wider variety of addresses.  Of the four `other' addresses, one of them 
belonged to ShapeShift, and the others belong to untagged clusters.

\begin{figure*}[t]
\centering
\begin{subfigure}[b]{0.32\textwidth}
\centering
\includegraphics[width=\linewidth]{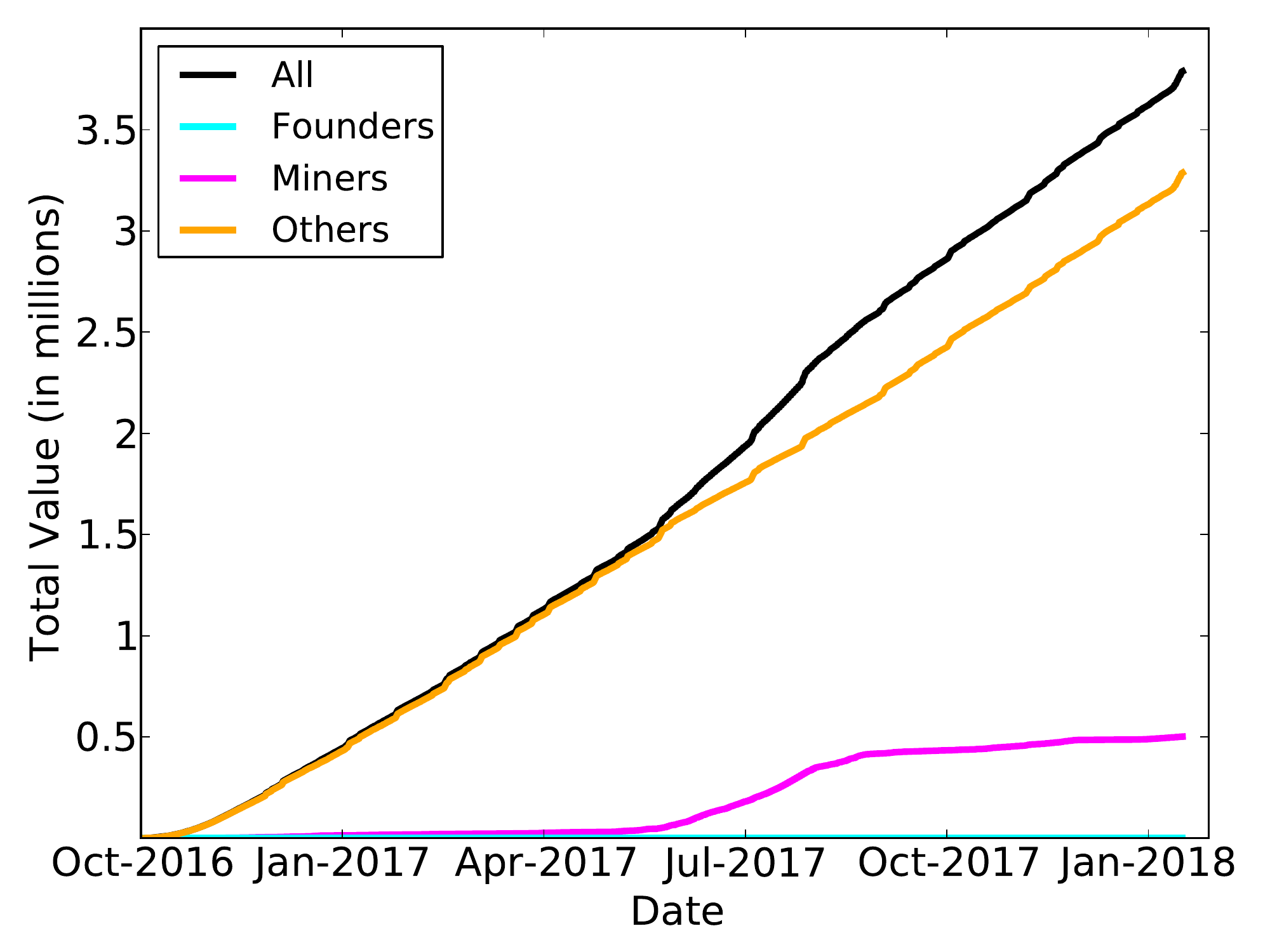}
\caption{No heuristics}
\label{fig:zt-none}
\end{subfigure}
~
\begin{subfigure}[b]{0.32\textwidth}
\includegraphics[width=\linewidth]{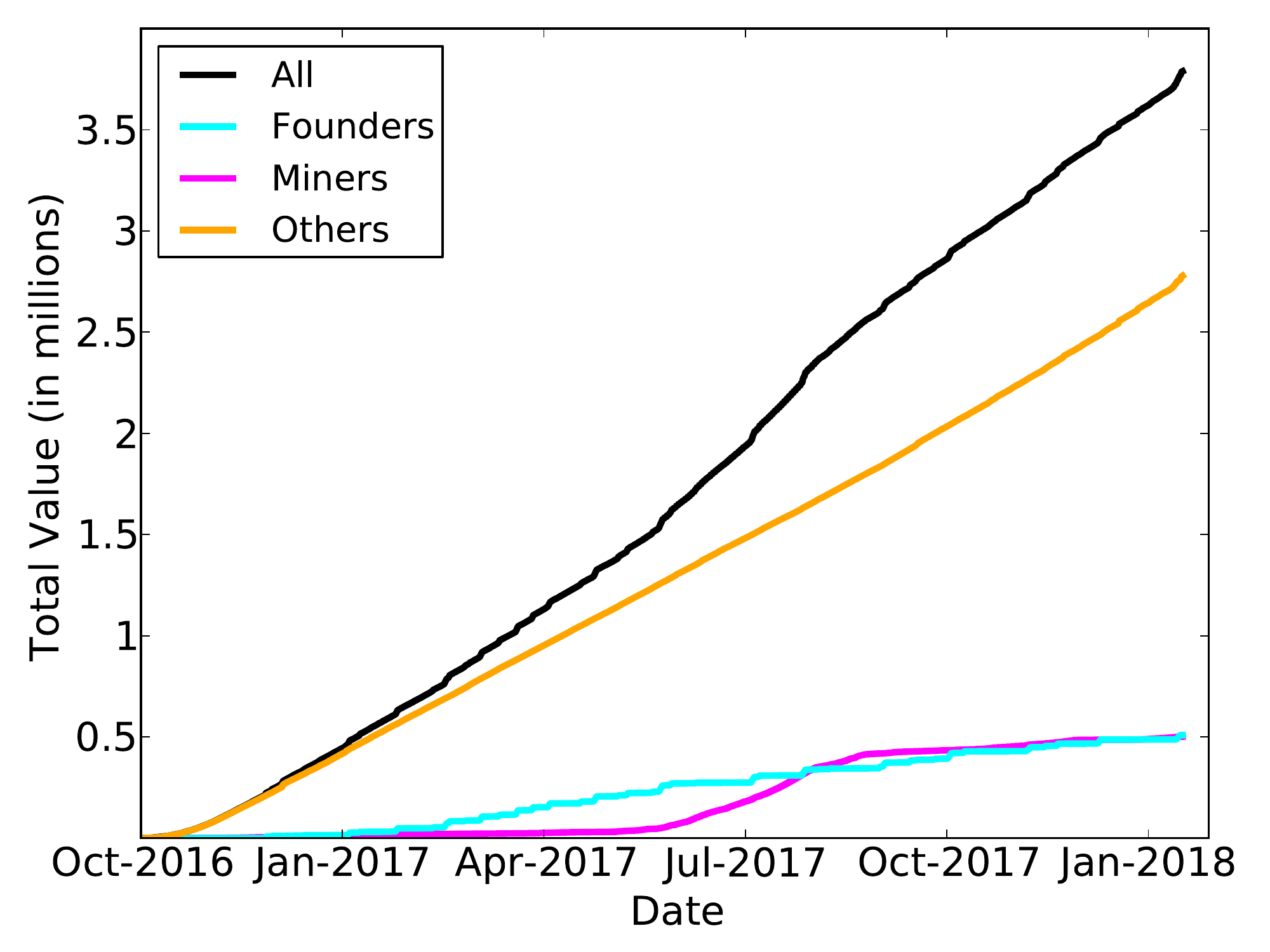}
\caption{Founder heuristic}
\label{fig:zt-founders}
\end{subfigure}
~
\begin{subfigure}[b]{0.32\textwidth}
\includegraphics[width=\linewidth]{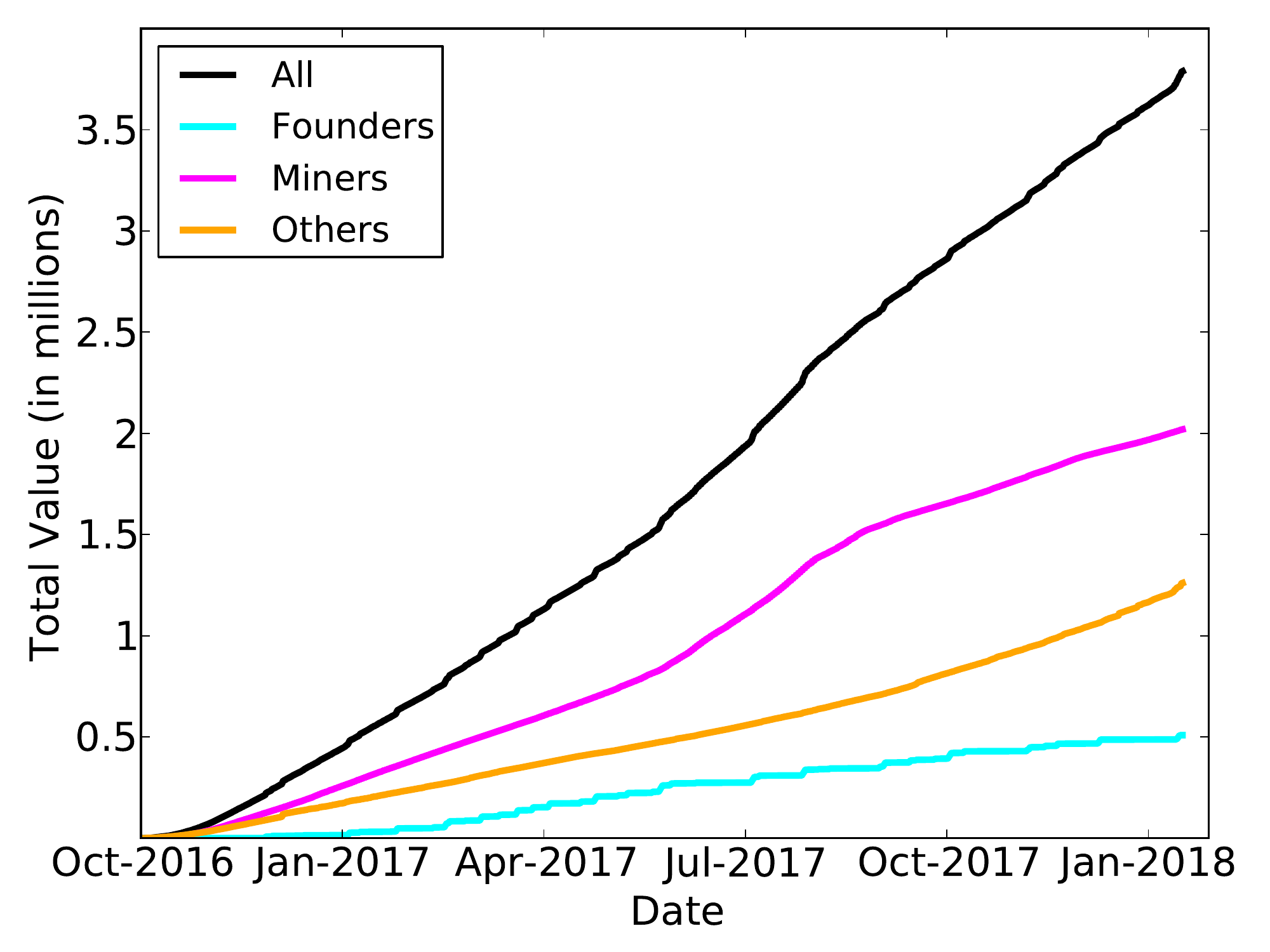}
\caption{Founder and miner heuristics}
\label{fig:zt-both}
\end{subfigure}
\caption{The \ztot transactions we associated with miners, founders, and
`other', after running some combination of our heuristics.}
\label{fig:all-heuristics}
\end{figure*}

While it is interesting to look at \ttoz transactions on their own, the main
intention of the shielded pool is to provide an anonymity set, so that when
users withdraw their coins it is not clear whose coins they are.  In that
sense, it is much more interesting to link together 
\ttoz and \ztot transactions, which acts to reduce the anonymity set.  More
concretely, if a \ttoz transaction can be linked to a \ztot transaction, then
those coins can be ``ruled out'' of the anonymity set of future users 
withdrawing coins from the pool.  
We thus devote our attention to this type of analysis for the rest of the
section.

The most na{\"i}ve way to link together these transactions would be to see if
the same addresses are used across them; i.e., if a miner uses the same
address to withdraw their coins as it did to deposit them.  By running this
simple form of linking, we see the results in Figure~\ref{fig:zt-none}.
This figure shows that we are not able to identify any withdrawals as
being associated with founders, and only a fairly small number as associated
with miners: 49,280 transactions in total, which account for 13.3\% of the
total value in the pool.  

Nevertheless, using heuristics that we develop for identifying founders (as
detailed in Section~\ref{sec:founders}) and miners (Section~\ref{sec:miners}),
we are able to positively link most of the \ztot activity with one of
these two categories, as seen in Figures~\ref{fig:zt-founders}
and~\ref{fig:zt-both}.  In the end, of the 177,009 \ztot transactions, we were 
able to tag 120,629 (or 68\%) of them as being associated with miners, 
capturing 52.1\% of the value coming out of the pool, and 2,103 of them as 
being associated with founders (capturing 13.5\% of the value).  We then
examine the remaining 30-35\% of the activity surrounding the shielded pool 
in Section~\ref{sec:others}.

\subsection{Founders}\label{sec:founders}

After comparing the list of founder addresses against the outputs of all 
coingen transactions, we found that 14 of them had been used.  Using 
these addresses, we were able to identify founder deposits into the pool, as 
already shown in Figure~\ref{fig:t-to-z}.  Table~\ref{tab:founders} provides a
closer inspection of the usage of each of these addresses.

\begin{table}[t]
\centering
\begin{tabular}{c S[table-format=4.0]S[table-format=6.1] S[table-format=3.0]} 
\toprule
& {\# Deposits} & {Total value} & {\# Deposits (249)} \\
\midrule
1 & 548 & 19600.4 & 0 \\ 
2 & 252 & 43944.6 & 153 \\
3 & 178 & 44272.5 & 177 \\
4 & 192 & 44272.5 & 176 \\
5 & 178 & 44272.5 & 177 \\ 
6 & 178 & 44272.5 & 177 \\
7 & 178 & 44272.5 & 177 \\
8 & 178 & 44272.5 & 177 \\
9 & 190 & 44272.5 & 176 \\ 
10 & 188 & 44272.5 & 176 \\
11 & 190 & 44272.5 & 176 \\
12 & 178 & 44272.5 & 177 \\
13 & 191 & 44272.5 & 175 \\
14 & 70 & 17500 & 70 \\
\midrule
Total & 2889 & 568042.5 & 2164 \\
\bottomrule
\end{tabular}
\caption{The behavior of each of the 14 active founder addresses, in terms of 
the number of deposits into the pool, the total value deposited (in ZEC), and 
the number of deposits carrying exactly 249.9999~ZEC in value.}
\label{tab:founders}
\end{table}

This table shows some quite obvious patterns in the behavior
of the founders.  At any given time, only one address is ``active,'' meaning it
receives rewards and deposits them into the pool.  Once it reaches the limit
of 44,272.5~ZEC, the next address takes its place and it is not used again.
This pattern has held from the third address onwards.  What's more, the amount
deposited was often the same: exactly 249.9999~ZEC, which is roughly the
reward for 100 blocks.  This was true of 74.9\%
of all founder deposits, and 96.2\% of all deposits from the third address
onwards.  There were only ever five other deposits into the pool carrying 
value between 249 and 251~ZEC (i.e., carrying a value close but not equal to
249.9999~ZEC).
%

Thus, while we were initially unable to identify any withdrawals associated
with the founders (as seen in Figure~\ref{fig:zt-none}), these patterns
indicated an automated use of the shielded pool that might also carry into 
the withdrawals.  Upon examining the withdrawals from the 
pool, we did not find any with a value exactly equal to 249.9999~ZEC.  We did,
however, find 1,953 withdrawals of exactly 250.0001~ZEC (and 1,969 carrying a
value between 249 and 251~ZEC, although we excluded the extra ones from our
analysis).

\begin{figure}[t]
\centering
\includegraphics[width=0.9\linewidth]{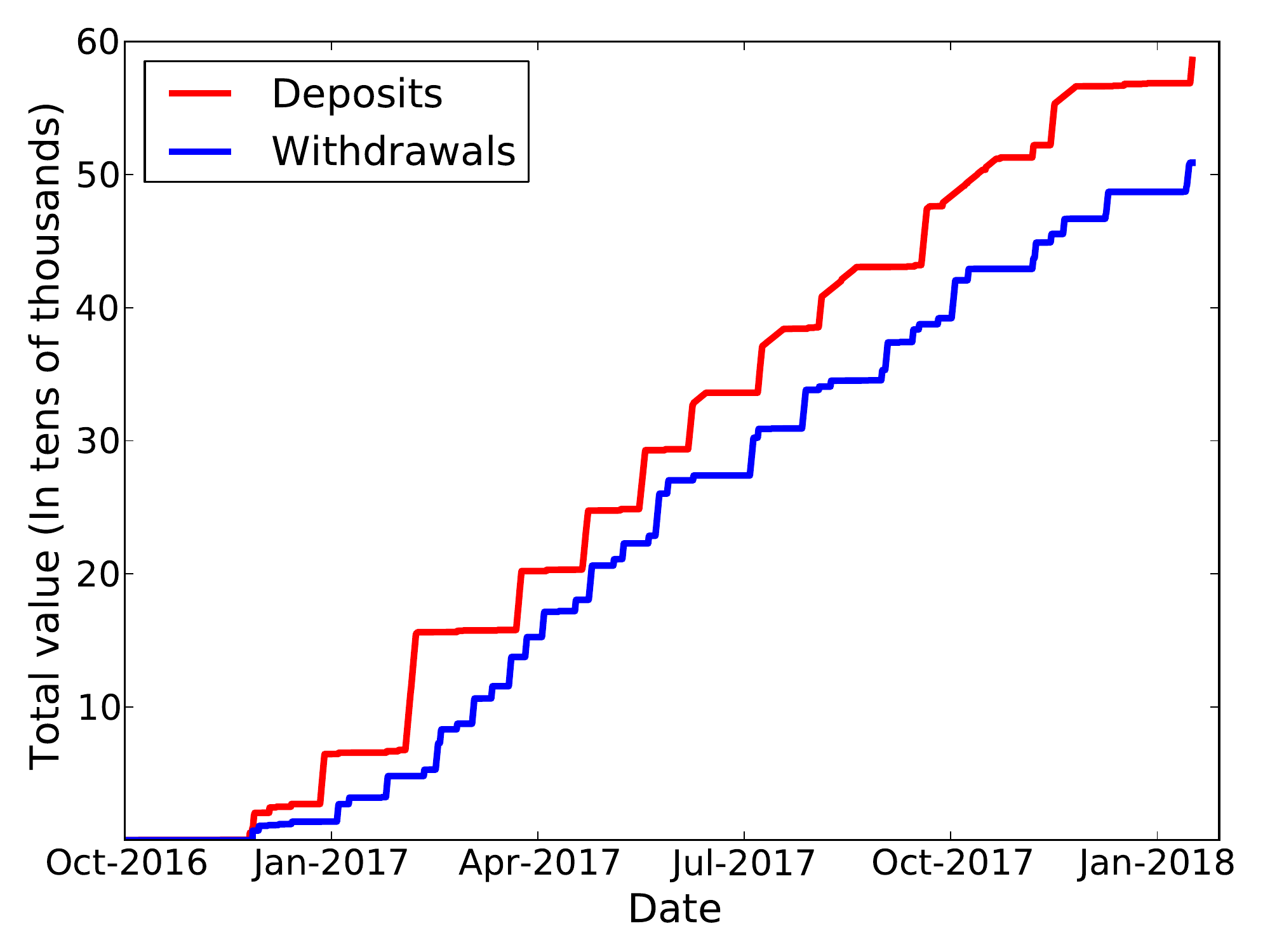}
\caption{Over time, the founder deposits into the pool (in red) and 
withdrawals from the pool (in blue), after running
Heuristic~\ref{heuristic:founder}.}
\label{fig:founders_correlation}
\end{figure}

The value alone of these withdrawals thus provides some correlation with the
deposits, but to further explore it we also looked at the timing of
the transactions.  When we examined the intervals between consecutive
deposits of 249.9999~ZEC, we found that 85\% happened within 
6-10 blocks of the previous one.  Similarly, when examining the intervals between
consecutive withdrawals of 250.0001~ZEC, we found that 1,943 of the 1,953
withdrawals also had a proximity of 6-10 blocks.  Indeed, both the deposits
and the withdrawals 
proceeded in step-like patterns, in which many transactions were made within 
a very small number of blocks (resulting in the step up), at which point 
there would be a pause while more block rewards were accumulated (the step
across).  This pattern is visible in Figure~\ref{fig:founders_correlation},
which shows the deposit and withdrawal transactions associated with
the founders.  Deposits are typically made in
few large steps, whereas withdrawals take many smaller ones. 

\begin{heuristic}\label{heuristic:founder}
Any \ztot transaction carrying 250.0001 ZEC in value is done by the founders.
\end{heuristic}

In terms of false positives, we cannot truly know how risky this
heuristic is, short of asking the founders.  This is in contrast to the
t-address clustering heuristics presented in Section~\ref{sec:T-to-T}, in
which we were not attempting to assign addresses to a specific owner, so 
could validate the heuristics in other ways.  Nevertheless, the high
correlation between both the value and timing of the transactions led us to
believe in the reliability of this heuristic.

As a result of running this heuristic, we added 75 more addresses to our
initial list of 48 founder addresses (of which, again, only 14 had been used).
Aside from the correlation showed in Figure~\ref{fig:founders_correlation}, 
the difference in terms of our ability to tag founder withdrawals is seen 
in Figure~\ref{fig:zt-founders}.

\subsection{Miners}\label{sec:miners}

The Zcash protocol specifies that all newly generated coins are required to be
put into the shielded pool before they can be spent further.  As a result, we 
expect that a large quantity of the ZEC being deposited into the pool are from
addresses associated with miners.  

\subsubsection{Deposits}

\begin{figure}[t]
\centering
\includegraphics[width=0.9\linewidth]{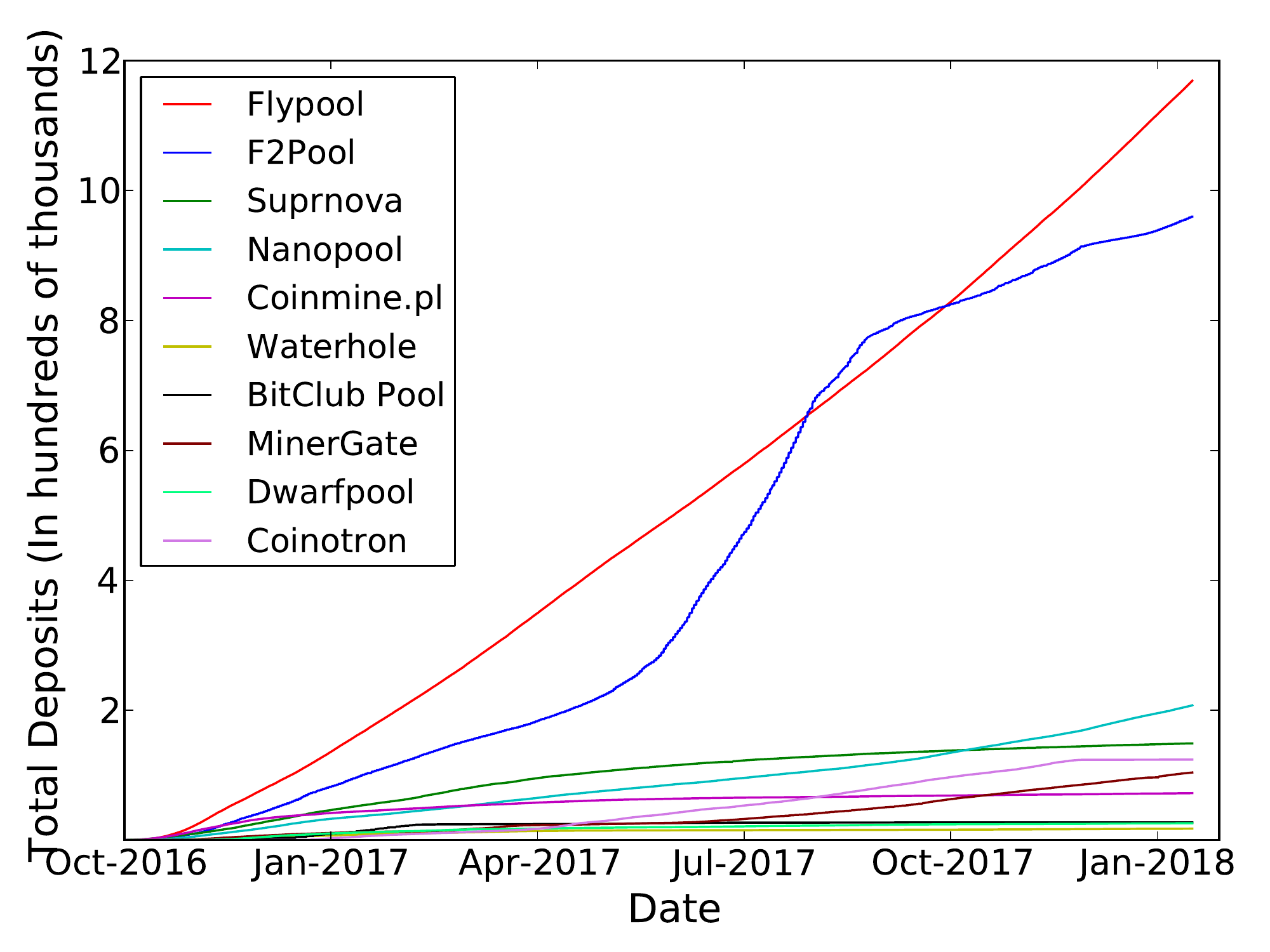}
\caption{Over time, the value of deposits made by known mining pools into the 
shielded pool.}
\label{fig:pools}
\end{figure}

As discussed earlier and seen in Figure~\ref{fig:t-to-z}, it is easy to identify 
miner deposits into the pool
due to the fact that they immediately follow a coin generation.  
%
Before going further, we split the category of miners into individual
miners, who operate on their own, and mining pools, which represent
collectives of potentially many individuals.  In total, we gathered 19 
t-addresses associated with Zcash mining pools, using the scraping methods
described in Section~\ref{sec:tagging}.  Table~\ref{tab:miners} lists these 
mining pools, as well as the number of addresses they control and the number 
of \ttoz transactions we associated with them.  Figure~\ref{fig:pools} plots 
the value of their deposits into the shielded pool over time.

\begin{table}
\centering
\begin{tabular}{lS[table-format=1.0]S[table-format=5.0]S[table-format=4.0]} 
\toprule
Name & {Addresses} & {\ttoz} & {\ztot} \\
\midrule
Flypool & 3 & 65631 & 3 \\ 
F2Pool & 1 & 742 & 720 \\
Nanopool & 2 & 8319 & 4107 \\
Suprnova & 1 & 13361 & 0 \\
Coinmine.pl & 2 & 3211 & 0 \\
Waterhole & 1 & 1439 & 5 \\
BitClub Pool & 1 & 196 & 1516 \\
MiningPoolHub & 1 & 2625 & 0 \\
Dwarfpool & 1 & 2416 & 1 \\
Slushpool & 1 & 941 & 0 \\
Coinotron & 2 & 9726 & 0 \\
Nicehash & 1 & 216 & 0 \\
MinerGate & 1 & 13 & 0 \\
Zecmine.pro & 1 & 6 & 0 \\ 
\bottomrule
\end{tabular}
\caption{A summary of our identified mining pool activity, in terms of the
number of associated addresses used in coingen transactions, and the numbers 
of each type of transaction interacting with the pool.}
\label{tab:miners}
\end{table}

In this figure, we can clearly see that the two dominant mining pools are
Flypool and F2Pool.  Flypool consistently
deposits the same (or similar) amounts, which we can see in their linear
representation.  F2Pool, on the other hand, has bursts of large deposits mixed
with periods during which it is not very active, which we can also see
reflected in the graph.  Despite their different behaviors, the amount
deposited between the two pools is similar.

\subsubsection{Withdrawals}

While the withdrawals from the pool do not solely re-use the small number
of mining addresses identified using deposits (as we saw in our na{\"i}ve 
attempt to link miner \ztot transactions in Figure~\ref{fig:zt-none}), they 
do typically re-use some of them, so can frequently be identified anyway.

In particular, mining pool payouts in Zcash are similar to how many of them 
are in Bitcoin~\cite{sarah-fistfulofbitcoins,SP:Eyal15}.  The block reward is 
often paid into a single address, controlled by
the operator of the pool, and the pool operator then deposits some set of 
aggregated block rewards into the shielded pool.  They then pay the individual 
reward to each of the individual miners as a way of ``sharing the pie,'' 
which results in \ztot transactions with many outputs. (In Bitcoin, some
pools opt for this approach while some form a ``peeling chain'' in which they
pay each individual miner in a separate transaction, sending the change back
to themselves each time.) In the payouts
for some of the mining pools, the list of output t-addresses sometimes 
includes one of the t-addresses known to be associated with the mining pool 
already.  We thus tag these types of payouts as belonging to the mining pool,
according to the following heuristic:

\begin{heuristic}\label{heuristic:miners}
If a \ztot transaction has over 100 output t-addresses, one of which belongs
to a known mining pool, then we label the transaction as a mining withdrawal
(associated with that pool), and label all non-pool output t-addresses as 
belonging to miners.
\end{heuristic}

As with Heuristic~\ref{heuristic:founder}, short of asking the mining pool
operators directly it is impossible to validate this heuristic.  Nevertheless,
given the known operating structure of Bitcoin mining pools and the way this
closely mirrors that structure, we again believe it to be relatively safe.

As a result of running this heuristic, we tagged 110,918 addresses as
belonging to miners, and linked a much more significant portion of the \ztot
transactions, as seen in Figure~\ref{fig:zt-both}.  As the last column in
Table~\ref{tab:miners} shows, however, this heuristic captured the activity 
of only a small number of the mining pools, and the large jump in linked
activity is mostly due to the high coverage with F2Pool (one of the two
richest pools).  This implies that further heuristics developed specifically
for other pools, such as Flypool, would increase the linkability even more.
Furthermore, a more active strategy in which we mined with the pools to
receive payouts would reveal their structure, at which point
(according to the 1.1M deposited by Flypool shown in Figure~\ref{fig:pools}
and the remaining value of 1.2M attributed to the `other' category shown in
Figure~\ref{fig:zt-both}) we would shrink the anonymity set even
further.\footnote{It is possible that we have already captured some of the
    Flypool activity, as many of the miners receive payouts from multiple pools.
    We thus are not claiming that all remaining activity could be attributed to
    Flypool, but potentially some substantial portion.}


\subsection{Other Entities}\label{sec:others}

Once the miners and founders have been identified, we can assume the remaining
transactions belong to more general entities.  In this section we look into 
different means of categorizing these entities in order to identify how the 
shielded pool is being used.

In particular, we ran the heuristic due to Quesnelle~\cite{zcash-anon}, which
said that if a unique value (i.e., a value never seen in the blockchain before
or since) is deposited into the pool and then, after some short
period of time, the exact same value is withdrawn from the pool, the deposit
and the withdrawal are linked in what he calls a 
\emph{round-trip transaction}.

\begin{heuristic}
{\textbf{\cite{zcash-anon}}}
\label{heuristic:bonus}
For a value $v$, if there exists exactly one \ttoz transaction carrying value 
$v$ and one \ztot transaction carrying value $v$, where the 
\ztot transaction happened after the \ttoz one and within some small number of
blocks, then these transactions are linked.
\end{heuristic}

In terms of false positives, the fact that the value is unique in the
blockchain means that the only possibility of a false positive is if some of
the \ztoz transactions split or aggregated coins in such a way that another 
deposit (or several other deposits) of a different amount were altered within 
the pool to yield an amount identical to the initial deposit.  While this is
possible in theory, we observe that of the 12,841 unique values we identified,
9,487 of them had eight decimal places (the maximum number 
in Zcash), and 98.9\% of them had more than three decimal
places.  We thus view it as highly unlikely that these exact values were
achieved via manipulations in \ztoz transactions.

\begin{figure}[t]
\centering
\includegraphics[width=0.9\linewidth]{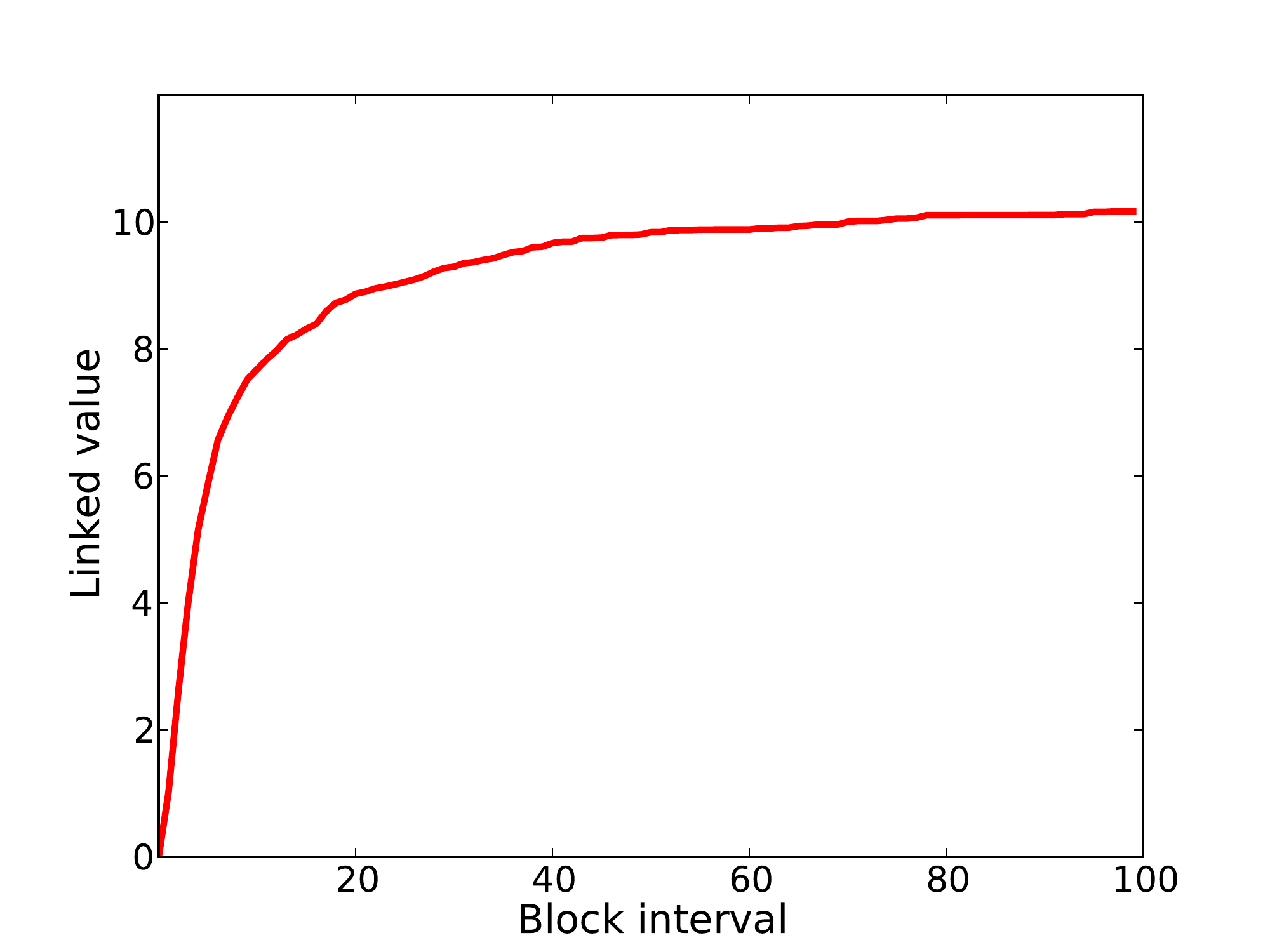}
\caption{The value linked by Heuristic~\ref{heuristic:bonus}, as a function of
the block interval required between the deposit and withdrawal transactions.}
\label{fig:bonus}
\end{figure}

By running this heuristic, we identified 12,841 unique values, which means 
we linked 12,841 transactions.  The values total 
1,094,513.23684~ZEC and represent 28.5\% of all coins 
ever deposited in the pool.  Interestingly, most (87\%) of the linked coins 
were in transactions attributed to the founders and miners, so had already 
been linked by our previous heuristics.  We believe this lends further 
credence to their soundness.  In terms of the block interval, we 
ran Heuristic~\ref{heuristic:bonus} for every interval between $1$ and $100$ 
blocks; the results are in Figure~\ref{fig:bonus}.

As this figure shows, even if we assume a conservative block interval of
$10$ (meaning the withdrawal took place 25 minutes after the deposit), we
still capture 70\% of the total value, or over 700K~ZEC.  If we require the
withdrawal to have taken place within an hour of the deposit, we get 83\%.

\section{Interactions within the Shielded Pool}\label{sec:z-to-z}


In this section we consider private transactions; i.e., \ztoz transactions
that interact solely with the shielded pool.  As seen in
Section~\ref{sec:tx-usage}, these transactions form a small percentage of the
overall transactions.  However, \ztoz transactions form a crucial part of the
anonymity core of Zcash.  In particular, they make it difficult to identify the
round-trip transactions from Heuristic~\ref{heuristic:bonus}.  

Our analysis identified 6,934 \ztoz transactions, with 8,444 \vjoinsplits.  As 
discussed in Section~\ref{sec:back-zcash}, the only information revealed by
\ztoz transactions is the miner's fee, the time of the transaction, and the 
number of \vjoinsplits used as input.  Of these, we looked at the time of
transactions and the number of \vjoinsplits in order to gain some insight as
to the use of these operations.

We found that 93\% of \ztoz transactions took just one \vjoinsplit as input.  
Since each \vjoinsplit can have at most two shielded outputs as its input, 
the majority of \ztoz transactions thus take no more than two shielded outputs 
as their input.  This increases the difficulty of categorizing \ztoz 
transactions, because we cannot know if a small number of users are making 
many transactions, or many users are making one transaction.

\begin{figure}[t]
\centering
\includegraphics[width=0.9\linewidth]{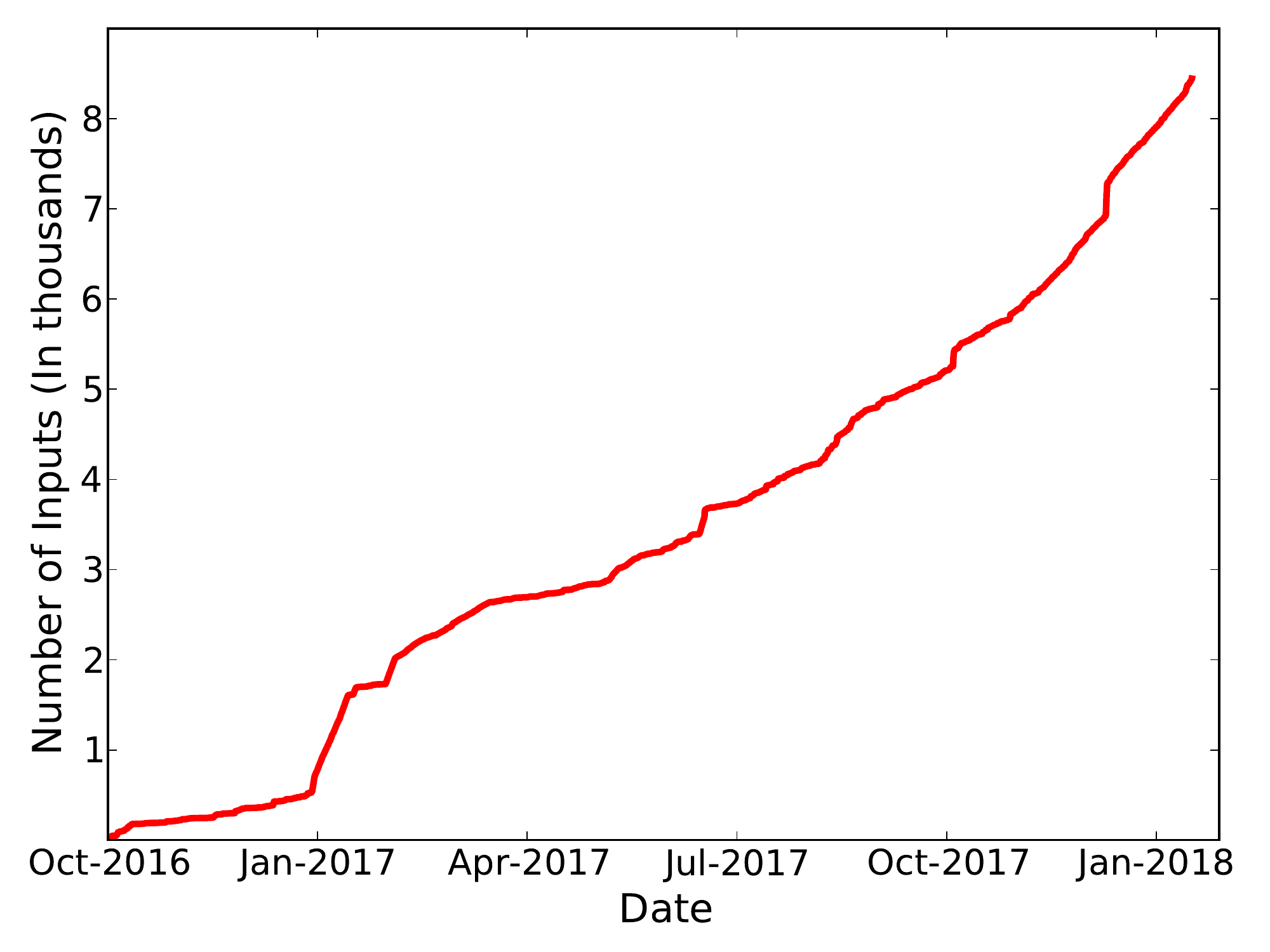}
\caption{The number of \ztoz \vjoinsplits over time.}
\label{fig:zz2}
\end{figure}

In looking at the timing of \ztoz transactions, however, we conclude that it 
is likely that a small number of users were making many transactions.
Figure~\ref{fig:zz2} plots the cumulative number of \vjoinsplits over time.
The occurrences of \vjoinsplits are somewhat irregular, with 17\%
of all \vjoinsplits occurring in January 2017.
There are four other occasions when a sufficient number of \vjoinsplits occur
within a sufficiently short period of time as to be visibly noticeable.  It 
seems likely that these occurrences belong to the same group of users, or at
least by users interacting with the same service.

Finally, looking back at the number of \ttoz and \ztot transactions identified
with mining pools in Table~\ref{tab:miners}, it is possible that BitClub Pool
is responsible for up to 1,300 of the \ztoz transactions, as it had 196
deposits into the pool and 1,516 withdrawals.  This can happen only because
either (1) the pool made extra \ztoz transactions, or (2) it sent change from 
its \ztot
transactions back into the shielded pool.  As most of BitClub Pool's \ztot
transactions had over 200 output t-addresses, however, we conclude that the
former explanation is more likely.

\section{Case Study: The Shadow Brokers}\label{sec:shadow-brokers}

The Shadow Brokers (TSB) are a hacker collective that has been active since
the summer of 2016, and that leaks tools supposedly created by the
NSA.  Some of these leaks are released as free samples, but many are sold
via auctions and as monthly bundles.  Initially, TSB accepted payment only 
using Bitcoin.  Later, however, they began to
accept Zcash for their monthly dump service.  In this section we discuss
how we identified \ttoz transactions that could represent payments to TSB.
We identified twenty-four clusters (created using our analysis in
Section~\ref{sec:T-to-T}) matching our criteria for potential TSB 
customers, one of which could be a regular customer.

\subsection{Techniques}

In order to identify the transactions that are most likely to be associated
with TSB, we started by looking at their blog~\cite{TSB}.  In May 2017, 
TSB announced that they would be accepting Zcash for their monthly dump 
service.
Throughout the summer (June through August) 
they accepted both Zcash and Monero, but in September they announced that they 
would accept only Zcash.
Table~\ref{table:TSB-blog-amounts} summarizes the amount they were requesting
in each of these months.  The last blog post was made in October 2017, when 
they stated that all subsequent dumps would cost 500~ZEC.

\begin{table}
\centering
\small
\begin{tabular}{ccccc}
\toprule
May/June & July & August & September & October \\
\midrule
\begin{tabular}[t]{l}
100 
\end{tabular}
&
\begin{tabular}[t]{l}
200 \\
400 \\
\end{tabular}
&
\begin{tabular}[t]{l}
500 \\ 
\end{tabular}
&
\begin{tabular}[t]{l}
100 \\
200 \\
500  
\end{tabular}
&
\begin{tabular}[t]{l}
500 \\
\end{tabular}
\\
\bottomrule
\end{tabular}
\caption{Amounts charged for TSB monthly dumps, in ZEC. In July and September 
TSB offered different prices depending on which exploits were being purchased.}
\label{table:TSB-blog-amounts}
\end{table}

To identify potential TSB transactions, we thus looked at all \ttoz 
transactions not associated with miners or founders that deposited either 
100, 200, 400, or 500 ZEC $\pm~5$ ZEC.  
Our assumption was that users paying TSB were not likely to be regular Zcash
users, but rather were using it with the main purpose of making the payment.  
On this basis, addresses making \ttoz transactions of the above values were 
flagged as a potential TSB customer if the following conditions held:

\begin{enumerate}

\item They did not get their funds from the pool; i.e., there were no \ztot
transactions with this address as an output.  Again, if this were a user 
mainly engaging with Zcash as a way to pay TSB, they would need to to buy
their funds from an exchange, which engage only with t-addresses.

\item They were not a frequent user, in the sense that they had not made 
or received more than 250 transactions (ever).

\item In the larger cluster in which this address belonged, the total amount
deposited by the entire cluster into the pool within one month was 
within 1~ZEC of the amounts
requested by TSB.  Here, because the resulting clusters were small enough to
treat manually, we applied not only Heuristic~\ref{heuristic:inputclusters}
but also Heuristic~\ref{heuristic:others} (clustering by change), making sure
to weed out false positives.   Again, the idea was that suspected TSB customers
would not be frequent users of the pool.

\end{enumerate}

As with our previous heuristics, there is no way to quantify the
false-positive risks associated with this set of criteria, although we see
below that many of the transactions matching it did occur in the time period
associated with TSB acceptance of Zcash.  Regardless, given this limitation we
are not claiming that our results are definitive, but do believe this 
to be a realistic set of criteria that might be applied in the
context of a law enforcement investigation attempting to narrow down potential
suspects.  

\subsection{Results}

Our results, in terms of the number of transactions matching our requirements
above up until 17 January 2018, are summarized in 
Table~\ref{table:number-of-suspicious-transactions-per-month}.  Before the
first TSB blog post in May, we found only a single matching transaction.  
This is very likely a false positive, but demonstrates that the types of 
transactions we were seeking were not common
before TSB went live with Zcash.  After the blog post, we flagged five clusters
in May and June for the requested amount of 100~ZEC.    There were only two 
clusters that was flagged for 500~ZEC, one of which was from August. 
No transactions of any of the required quantities were flagged in September,
despite the fact that TSB switched to accepting only Zcash in September.  This 
is possible for a number of reasons: our criteria may have caused us to miss 
transactions, or maybe there were no takers.  From October onwards we flagged 
between 1-6 transactions per month.  It is hard to know if these represent users 
paying for old data dumps or are simply false positives.  

\begin{table}
\centering
\begin{tabular}{lcccc}
\toprule
Month & 100 & 200 & 400 & 500 \\
\midrule
October (2016)   & 0   &  0  &  0  &  0  \\
November         & 0   &  0  &  0  &  0  \\
December         & 0   &  0  &  0  &  0  \\
January (2017)   & 1   &  0  &  0  &  0  \\
February         & 0   &  0  &  0  &  0  \\
March            & 0   &  0  &  0  &  0  \\
April            & 0   &  0  &  0  &  0  \\
May (before) & 0   &  0  &  0  &  0  \\
May (after)  & 3   &  1  &  0  &  0  \\
June       & 2   &  1  &  1  &  0  \\
July       & 1   &  2  &  0  &  0  \\
August     & 1   &  0  &  0  &  1  \\
September  & 0   &  0  &  0  &  0  \\
October    & 2   &  0  &  0  &  0  \\
November   & 1   &  0  &  0  &  0  \\
December   & 2   &  3  &  0  &  1  \\
January (2018)   & 0   &  1  &  0  &  0 \\
\bottomrule
\end{tabular}
\caption{Number of clusters that put the required amounts ($\pm 1$~ZEC) 
into the shielded pool.}
\label{table:number-of-suspicious-transactions-per-month}
\end{table}

Four out of the 24 transactions in 
Table~\ref{table:number-of-suspicious-transactions-per-month} are highly 
likely to be false positives.
First, there is the deposit of 100~ZEC into the pool in January, before
TSB announced their first blog post.  This cluster put an additional 252~ZEC
into the pool in March, so is likely just some user of the pool.
Second and third, there are two deposits of 200~ZEC into the pool in 
June, before TSB announced that one of the July dump prices would cost 
200~ZEC.  
Finally, there is a deposit of 400~ZEC into the pool in June before TSB 
announced that one of the July dump prices would cost 400~ZEC.

Of the remaining clusters, there is one whose activity is worth discussing.
From this cluster, there was one deposit into the pool in
June for 100~ZEC, one in July for 200~ZEC, and one in August for 500~ZEC,
matching TSB prices exactly.  The cluster belonged to a new user, and most of
the money in this user's cluster came directly from Bitfinex (Cluster~3).


\section{Conclusions}\label{sec:conclusions}

This paper has provided the first in-depth exploration of Zcash, with a
particular focus on its anonymity guarantees.  To achieve this, we applied
both well-known clustering heuristics that have been developed for Bitcoin
and attribution heuristics we developed ourselves that take into account 
Zcash's shielded pool and its unique cast of characters.  As with previous 
empirical analyses of other cryptocurrencies, our study has shown that most 
users are not taking advantage of the main privacy feature of Zcash at all.
Furthermore, the participants who do engage with the shielded pool do so in a
way that is identifiable, which has the effect of significantly eroding the 
anonymity of other users by shrinking the overall anonymity set.  

\subsubsection*{Future work}  

Our study was an initial exploration, and thus left many avenues open for
further exploration.  For example, it may be possible to classify more \ztoz 
transactions by analyzing the time intervals between the transactions in more 
detail, or by examining other metadata such as the miner's fee or even the
size (in bytes) of the transaction.  Additionally, the
behavior of mining pools could be further identified by a study that actively
interacts with them.

\subsubsection*{Suggestions for improvement}

Our heuristics would have been significantly less effective if the founders
interacting with the pool behaved in a less regular fashion.  In particular,
by always withdrawing the same amount in the same time intervals, it became
possible to distinguish founders withdrawing funds from other users.  Given
that the founders are both highly invested in the currency and knowledgeable
about how to use it in a secure fashion, they are in the best place to ensure
the anonymity set is large.

Ultimately, the only way for Zcash to truly ensure the
size of its anonymity set is to require all transactions to take place within
the shielded pool, or otherwise significantly expand the usage of it.  This 
may soon be computationally feasible given emerging advances in the 
underlying cryptographic techniques~\cite{jubjub}, or even if more mainstream
wallet providers like Jaxx roll out support for z-addresses.  More broadly, 
we view it as an interesting 
regulatory question whether or not mainstream exchanges would continue to 
transact with Zcash if it switched to supporting only z-addresses.

\ifsubmission\else{
\section*{Acknowledgments}

We would like to thank Lustro, the maintainer of the Zchain explorer, for
answering specific questions we asked about the service.  
The authors are supported in part by EPSRC Grant EP/N028104/1, and in part by
the EU H2020 TITANIUM project under grant agreement number 740558.  Mary Maller
is also supported by a scholarship from Microsoft Research.

}\fi

{
\balance
{\footnotesize
\bibliographystyle{abbrv}
\bibliography{abbrev2,crypto,misc}
}
}


\end{document}